\pdfoutput=1
\documentclass[11pt]{article}
\usepackage{algorithm}
\usepackage{algpseudocode}
\usepackage{amsfonts}
\usepackage{amsmath}
\usepackage{amssymb}
\usepackage{array}
\usepackage{authblk}
\usepackage{booktabs}
\usepackage[margin=1in]{geometry}
\usepackage{graphicx}
\usepackage{setspace}
\usepackage{subcaption}
\usepackage{xcolor}
\usepackage[colorlinks=true,citecolor=blue]{hyperref}

\definecolor{relatedorange}{RGB}{191,87,0}

\graphicspath{ {./figures/} }

\newcommand{\splitcell}[1]{\begin{tabular}[t]{@{}l@{}}#1\end{tabular}}
\newcommand{\goodcell}[1]{\begingroup\color{green!55!black}#1\endgroup}
\newcommand{\relatedcell}[1]{\begingroup\color{relatedorange}#1\endgroup}
\newcommand{\badcell}[1]{\begingroup\color{red}#1\endgroup}

\title{FoundPDE: A generative foundation model for data-driven symbolic discovery of partial differential equations}

\author[1,$\dag$]{Langchen Liu}
\author[1,2,$\dag$]{Junlin Luo}
\author[1]{Min Zhu}
\author[1]{Zhuoran Yang}
\author[1,3,*]{Lu Lu}

\affil[1]{Department of Statistics and Data Science, Yale University, New Haven, CT 06511, USA}
\affil[2]{Department of Mechanical Engineering, Stanford University, Stanford, CA 94305, USA}
\affil[3]{Department of Chemical and Environmental Engineering, Yale University, New Haven, CT 06511, USA}
\affil[*]{Corresponding author. Email: lu.lu@yale.edu}
\affil[$\dag$]{These authors contributed equally to this work.}
\date{}

\begin{document}

\maketitle

\begin{abstract}
Discovering governing partial differential equations (PDEs) from observed data is a central inverse problem in computational science and engineering: the goal is not only to predict trajectories, but to identify interpretable symbolic laws that can be analyzed and reused. Existing PDE discovery methods often depend on hand-designed libraries, accurate numerical derivatives, or per-equation symbolic search, making them fragile under sparse observations, noisy data, broad equation families, and unseen symbolic compositions. Here, we introduce FoundPDE, a pretrained data-to-symbol foundation model for PDE discovery. FoundPDE maps discretized spatio-temporal trajectories to symbolic equations with a transformer, uses Fourier and first-order spatial-derivative augmentation to expose complementary trajectory information, refines numerical coefficients, and supports downstream extension with improved generalization through few-shot fine-tuning and hierarchical discovery. On a clean in-domain benchmark spanning 20 PDE families, our model reaches \(97.50\%\) type recognition. Under coefficient shifts beyond the training support, type recognition remains \(90.53\%\); under multiplicative Gaussian noise on the model input, it is \(94.43\%\) at a \(2\%\) noise level but falls to \(52.09\%\) at \(10\%\). For unseen composite equations, few-shot fine-tuning reaches over \(90\%\) exact recovery on four held-out families, and we further develop a hierarchical discovery algorithm that recovers two-term composites without fine-tuning FoundPDE. By amortizing symbolic discovery across equation families, FoundPDE shifts PDE discovery from per-dataset search toward a reusable pretrained model that extends to new physics with limited additional data, a step toward foundation models that uncover interpretable governing laws directly from scientific data.
\end{abstract}


\section{Introduction}
Partial differential equations (PDEs) are central to modeling transport, diffusion, wave propagation, reaction dynamics, and many other physical processes \cite{karniadakis2021physics}. Recovering a PDE directly from observational data is therefore a scientifically important inverse problem: beyond prediction, it provides an interpretable law that can be analyzed, validated, and transferred across related regimes. Recent advances in scientific machine learning (SciML) have made data-driven PDE discovery increasingly feasible, but robust symbolic PDE discovery remains challenging when observations are sparse, noisy, or drawn from many possible equation families \cite{osti_1478744}.

SciML integrates physical principles with machine learning to yield models that adhere closely to scientific laws. Physics-Informed Neural Networks (PINNs)~\cite{raissi2019physics,lu2021deepxde} incorporate PDE constraints directly into their training, enhancing interpretability and data efficiency. Operator learning frameworks, including Deep Operator Networks (DeepONets)~\cite{lu2019deeponet, deng2022approximation, lin2021operator} and Fourier Neural Operators (FNOs)~\cite{li2021fourierneuraloperatorparametric}, have further expanded capabilities by efficiently learning mappings between function spaces. Recent hybrid models such as Physics-Informed Neural Operators (PINOs)~\cite{li2023physicsinformedneuraloperatorlearning} and physics-informed DeepONets~\cite{wang2021learningsolutionoperatorparametric} combine these methods, addressing complex multi-scale phenomena and improving generalization.

Building on this broader SciML progress, foundation models have significantly advanced scientific modeling by adopting large-scale pretraining techniques from natural language processing. Models such as ICON~\cite{Yang_2023} leverage in-context learning, enabling zero-shot generalization across diverse PDEs. Multimodal architectures such as PROSE~\cite{LIU2024106707} and PROSE-PDE~\cite{sun2025foundationmodelpartialdifferential} integrate symbolic knowledge with learned representations, effectively generalizing across multiple physics scenarios. Similarly, PDEformer~\cite{ye2025pdeformerfoundationmodelonedimensional} uses transformer-based computational graph representations to provide mesh-free, zero-shot predictions across varied PDEs, demonstrating the extensive potential of pre-trained models for scalable and versatile PDE solutions. PI-MFM~\cite{zhu2025pimfm} further enforces the governing equations during pretraining and adaptation by assembling physics-informed residual losses directly from the symbolic PDE input.
These developments show that pretrained scientific models can be reused across families of PDE-related tasks, but most of this progress has focused on solution prediction or operator learning. The present paper instead targets symbolic discovery: the goal is not only to predict a future trajectory, but to recover an explicit governing equation from observed solution data.

Symbolic PDE discovery aims to recover explicit governing equations from data~\cite{lou2026datadrivendiscovery}. Sparse-regression methods such as SINDy~\cite{Brunton_2016} and PDE-FIND~\cite{doi:10.1126/sciadv.1602614} select active terms from predefined candidate libraries. They are efficient and interpretable when the library contains the right terms and derivatives can be estimated accurately, but become fragile when data are low-resolution or noisy, or when the equation contains terms outside the library. UPINN~\cite{podina2023upinn} and D-CIPHER~\cite{kacprzyk2023dcipher} address sparse-data and derivative-mismatch regimes, while search-based, physics-constrained, and prior-informed approaches, including Symbolic Physics Learner~\cite{sun2023symbolicphysicslearner}, DISCOVER~\cite{PhysRevResearch.6.013182}, PhysPDE~\cite{feng2025physpde}, ABL-PDE~\cite{gao2025abductivepde}, and symmetry-informed discovery~\cite{yang2024symmetryinformed}, enlarge or constrain the symbolic hypothesis space at the price of higher computational cost and a separate optimization for every equation instance. In general symbolic regression, differentiable genetic programming relaxes discrete expression trees so that gradient-based optimization can scale tree search to high-dimensional inputs~\cite{song2026dgp}. Transformer-based symbolic models~\cite{lample2019deeplearningsymbolicmathematics,kamienny2022endtoendsymbolicregressiontransformers}, ODEFormer~\cite{dascoli2023odeformersymbolicregressiondynamical}, amortized equation discovery~\cite{liu2024amortizedequation}, and methods based on large language models (LLMs), such as LLM-SR~\cite{shojaee2025llmsr} and LLM-SRBench~\cite{shojaee2025llmsrbench}, instead treat equation discovery as a reusable prediction or program-generation problem, but existing approaches for PDE discovery remain limited in scope or rely on additional symbolic priors and stage-wise procedures.

Together, these limitations motivate a foundation-model approach to symbolic PDE discovery. In this paper, we use the term \emph{foundation model} in a local and operational sense: a model pretrained on a broad, canonical collection of one-dimensional time-dependent PDE families and reused across downstream symbolic discovery tasks rather than optimized for one equation instance. The claim rests on three ingredients: coverage of representative PDE structures during pretraining, reuse of one pretrained model across many discovery tasks, and evidence that the learned representation supports parameter generalization and targeted downstream adaptation. Our goal is not to build a universal PDE solver or to claim exhaustive coverage of one-dimensional dynamics. Instead, FoundPDE targets symbolic discovery: it takes discretized spatio-temporal solution data as input and predicts a symbolic expression for the governing PDE. This framing separates the present work from operator-learning models that predict future solution fields; here, the output is an interpretable equation.

We propose \textbf{FoundPDE}, a transformer-based generative foundation model that casts PDE discovery as data-to-symbol sequence generation. The model encodes solution snapshots, optionally augmented with spatial Fourier features and first-order spatial derivative features, and autoregressively decodes a tokenized symbolic PDE representation. A post-processing step then refines the predicted floating-point coefficients without changing the PDE symbolic structure. We evaluate parameter out-of-distribution (OOD) generalization by testing whether PDE families remain recognizable on coefficient supports that are strictly disjoint from the in-domain ranges. We then study few-shot fine-tuning and develop hierarchical discovery as two preliminary routes for extending the model beyond the direct pretraining vocabulary.

Our contributions are as follows. First, we develop a foundation model for direct symbolic discovery of time-dependent one-dimensional PDEs from solution data. Second, we construct a large synthetic training corpus spanning 20 representative PDE families and study how Fourier and derivative augmentations affect type recognition rate (TRR), mean relative coefficient squared error (MRCSE), relative \(L^2\) error, and training cost. Third, we evaluate robustness to multiplicative noise on the model input in sparse-data settings. Finally, we evaluate parameter-OOD generalization, few-shot fine-tuning, and hierarchical discovery, clarifying both the promise and the current limitations of a foundation-model view of symbolic PDE discovery.

\section{Methods}
\label{sec:methodology}

This section defines the symbolic PDE discovery problem and describes the main components of FoundPDE. Implementation-level architecture, training, augmentation, dataset-generation, and hardware details are deferred to Appendixes~\ref{app:dataset_specification} and \ref{app:model_training_details}, so the main text focuses on the discovery formulation, coefficient refinement, evaluation metrics, and extension protocols.

\subsection{Problem setup}
\label{sec:problem_setup}

We consider one-dimensional time-dependent PDEs whose solutions are observed on a spatio-temporal grid. For each instance, the available data are a trajectory
\[
\mathbf{u}=\{u(t_i,x_j)\}_{i=1,j=1}^{N_t,N_x}\in\mathbb{R}^{N_t\times N_x},
\]
generated by an unknown governing equation. The goal is to recover the underlying symbolic PDE expression. In the experiments, the target equations can be written abstractly as
\[
\partial_t^{r_t}u
= \mathcal{G}(u,u_x,u_{xx},u_{xxx},u_{xxxx},x;\boldsymbol{\theta}),
\qquad r_t\in\{1,2\},
\]
where \(\mathcal{G}\) is a symbolic operator assembled from arithmetic operations, differential terms, elementary functions, and numerical coefficients \(\boldsymbol{\theta}\). The temporal order \(r_t\) is part of the symbolic target: first-order families use \(r_t=1\), whereas PDEs such as the wave, Klein--Gordon, and Sine--Gordon equations use \(r_t=2\). Accordingly, the output vocabulary contains both \(u_t\) and \(u_{tt}\).

Following prior works on symbolic mathematics and equation discovery~\cite{lample2019deeplearningsymbolicmathematics,kamienny2022endtoendsymbolicregressiontransformers,LIU2024106707}, we represent each symbolic PDE using Polish notation, i.e., a prefix traversal of the expression tree. This converts a structured symbolic object into a sequence that can be generated autoregressively while preserving the underlying operator hierarchy. The symbolic discovery task is formulated as conditional sequence modeling~\cite{vaswani2017attention,kamienny2022endtoendsymbolicregressiontransformers,dascoli2023odeformersymbolicregressiondynamical}:
\[
P(\tau \mid \mathbf{u})=\prod_{\ell=1}^{L}P(\tau_\ell\mid \tau_{<\ell},\mathbf{u}),
\]
where \(\tau=(\tau_1,\ldots,\tau_L)\) is the target token sequence. This setup fixes the central input-output object of the paper: FoundPDE maps numerical solution data to an interpretable governing equation. The model is trained to predict a tokenized representation of the complete equation, including its temporal derivative order, from the observed trajectory.

\subsection{FoundPDE framework}
\label{sec:framework}

\begin{figure}[htbp]
    \centering
    \includegraphics[width=\linewidth]{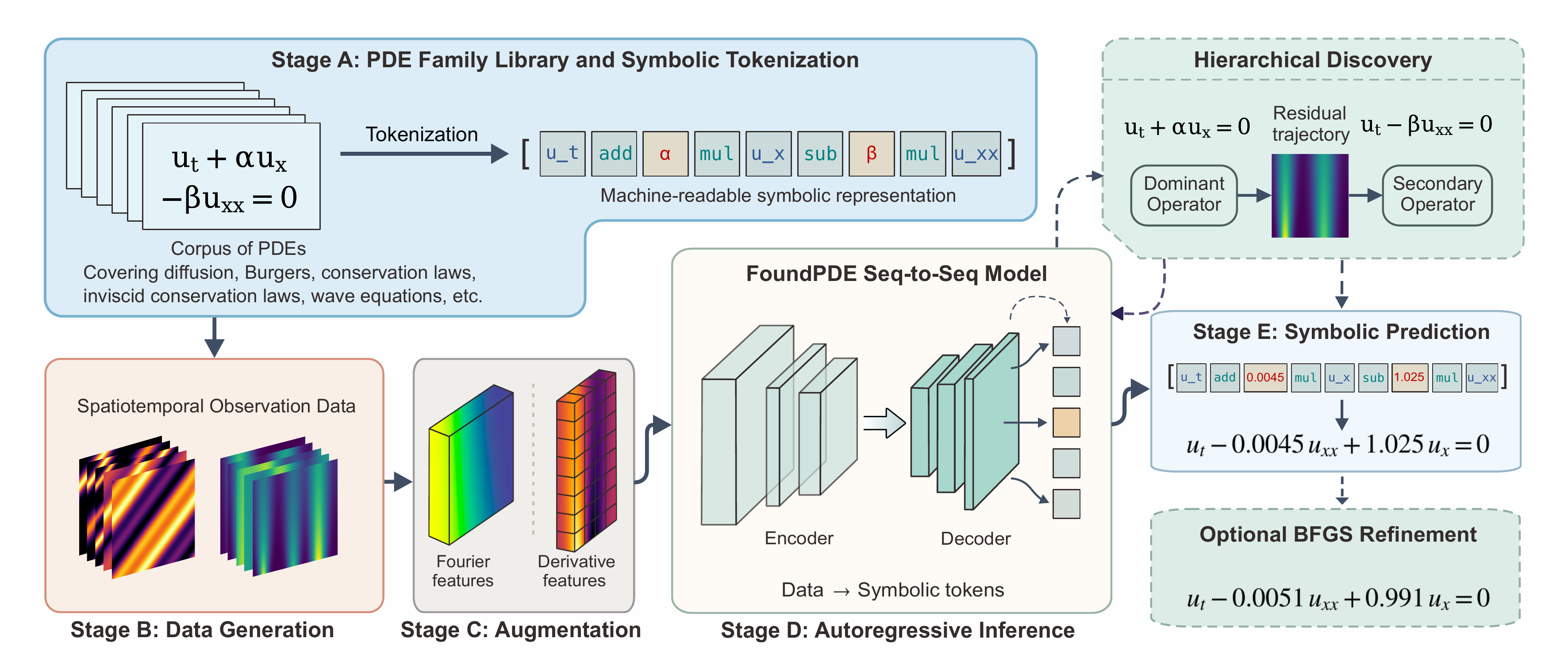}
    \caption{\textbf{FoundPDE performs end-to-end symbolic PDE discovery from discretized solution data.} Stage A constructs the symbolic training corpus and tokenizes each governing equation. Stage B solves the sampled equations under randomized initial conditions to produce spatio-temporal observations. Stage C optionally augments each trajectory with first-order spatial-derivative and Fourier channels. Stage D encodes these observations and autoregressively decodes a symbolic PDE in Polish notation. Stage E returns the predicted expression. The optional BFGS branch updates only its floating-point coefficients, whereas hierarchical discovery reuses the pretrained model to recover selected composite operators sequentially.}
    \label{fig:architecture}
\end{figure}

FoundPDE uses a transformer-based data-to-symbol architecture (Fig.~\ref{fig:architecture}) to amortize symbolic discovery across a pretrained corpus of PDE families~\cite{vaswani2017attention,LIU2024106707,dascoli2023odeformersymbolicregressiondynamical}. The encoder maps raw or augmented solution trajectories into latent representations, and the decoder autoregressively generates the symbolic PDE sequence. In the main experiments, we compare four input configurations: the raw solution field, the solution field with Fourier features, the solution field with a first-order spatial derivative channel, and the combination of Fourier and spatial-derivative channels. Full architecture and training settings are reported in Appendix~\ref{app:model_training_details}.

\paragraph{Data embedding and encoder.}
The raw and augmented input channels are first projected into a latent representation before being processed by the transformer encoder. Positional information is added so that the model can distinguish spatial and temporal locations in the observed trajectory. Self-attention allows the encoder to aggregate spatial and temporal dependencies across the full solution field, giving the decoder a global representation of the candidate dynamics rather than a pointwise derivative estimate alone~\cite{vaswani2017attention}.

\paragraph{Symbolic decoder.}
The decoder autoregressively generates a symbolic PDE in Polish notation. The output vocabulary contains arithmetic operators, differential operators, state-dependent functions, and numerical-constant tokens. Floating-point coefficients are represented by a tokenized numeric scheme, while the operator tokens determine the symbolic structure. Decoding proceeds until an end-of-sequence token is produced.

\paragraph{Normalization.}
We apply reversible instance normalization (RevIN)~\cite{kim2022revin} to stabilize training across PDE families with different solution scales. This normalization is useful because the corpus contains equations with distinct physical units, coefficient ranges, and trajectory amplitudes.

\subsection{Feature augmentation}
\label{sec:augmentation}

The raw solution field alone may not expose all structures needed for symbolic recovery, especially when different PDEs generate similar trajectories over a finite observation window. We therefore study two augmentation channels: Fourier features, which summarize global spatial-frequency content, and first-order spatial derivative features, which expose local variation in the solution field.

\paragraph{Global Fourier feature augmentation.}
For each solution trajectory, we compute the discrete Fourier transform along the spatial dimension. The real and imaginary parts of the transformed signal are concatenated with the original solution channel. These features are intended to help the encoder distinguish dynamics with characteristic oscillatory or dispersive behavior, such as wave-like or higher-order spatial operators.

\paragraph{Local spatial derivative augmentation.}
For derivative augmentation, we compute the first-order spatial derivative \(u_x\) using finite differences along the spatial axis. This channel is concatenated with the original solution field. The augmentation does not provide \(u_t\) or \(u_{xx}\) directly; it supplies only local first-order spatial variation. This design gives the model additional differential information while preserving the task of inferring the governing symbolic structure.

\paragraph{Feature concatenation.}
Depending on the experimental variant, the encoder input is one of four channel configurations: (1) the bare solution field \(u\), (2) \(u\) with Fourier features, (3) \(u\) with \(u_x\), or (4) \(u\) with both \(u_x\) and Fourier features. We evaluate these variants separately to determine whether global frequency information and local spatial-gradient information contribute differently to type recognition, coefficient estimation, and trajectory fidelity.

\subsection{PDE coefficient refinement}
\label{sec:bfgs}

The transformer predicts both symbolic structure and approximate numerical coefficients. Because coefficient tokens are discretized and generated autoregressively, their raw values can be inaccurate even when the operator structure is correct. We therefore apply BFGS~\cite{nocedal2006numerical} as a post-processing step that refines only the floating-point coefficients while keeping the predicted symbolic structure fixed.

Given a predicted structure \(\widehat{\mathcal{G}}\) and initial coefficient vector \(\boldsymbol{\theta}_0\), we solve
\[
\widehat{\boldsymbol{\theta}}
=\arg\min_{\boldsymbol{\theta}}
\frac{1}{K}\sum_{k=1}^{K}
\left\|
\widehat{\partial_t^{r_t}u}(t_k,x_k)
-
\widehat{\mathcal{G}}\bigl(u(t_k,x_k),\mathcal{D}_x u(t_k,x_k);\boldsymbol{\theta}\bigr)
\right\|_2^2.
\]
Here \(K\) is the number of residual-evaluation points and \(\mathcal{D}_x u=(u_x,u_{xx},u_{xxx},u_{xxxx})\) collects the supported spatial derivatives. The temporal derivative order \(r_t\) follows the decoded symbolic target, and \(\widehat{\partial_t^{r_t}u}\) and the required components of \(\mathcal{D}_x u\) are estimated from the observed trajectory. This refinement is numerical calibration, not a second symbolic search stage. It can improve numerical coefficients and relative \(L^2\) reconstruction, but it cannot add missing terms or remove spurious terms. Additional implementation details are given in Appendix~\ref{app:bfgs_details}.

\subsection{Evaluation metrics}
\label{sec:metrics}

We evaluate each prediction using three complementary metrics: type recognition rate (TRR), mean relative coefficient squared error (MRCSE), and relative \(L^2\) error. These metrics are reported separately because a prediction may recover the correct PDE family but assign inaccurate coefficients, or recover an incorrect symbolic type that still produces a similar solution over a finite time window.

\paragraph{Type recognition rate.}
Let \(\mathcal{T}(y_i)\) denote the PDE type of the ground-truth expression for sample \(i\), and let \(\mathcal{T}(\hat{y}_i)\) denote the type assigned to the parsed prediction. A prediction is counted as type-correct when
\[
\mathbf{1}_{\mathrm{type}}(i)=
\mathbf{1}\!\left[\mathcal{T}(\hat{y}_i)=\mathcal{T}(y_i)\right].
\]
The type recognition rate is
\[
\mathrm{TRR}
=
\frac{1}{N}\sum_{i=1}^{N}\mathbf{1}_{\mathrm{type}}(i).
\]
Type recognition is therefore a structural metric: it measures whether the predicted expression belongs to the correct canonical PDE family, not whether every displayed coefficient is numerically accurate.

\paragraph{Mean relative coefficient squared error.}
MRCSE is computed only on type-matched predictions for which the relevant coefficients can be extracted. Let \(\mathbf{c}_i\in\mathbb{R}^{d_i}\) be the ground-truth packed coefficient vector for sample \(i\), i.e., the physical parameters of its PDE family (e.g., the wave speed \(\beta\) rather than its displayed square \(\beta^2\)), and let \(\hat{\mathbf{c}}_i\) be the corresponding predicted coefficient vector. For the evaluable type-matched set \(\mathcal{I}_{c}\), we report
\[
\mathrm{MRCSE}
=
\frac{1}{|\mathcal{I}_{c}|}
\sum_{i\in\mathcal{I}_{c}}
\frac{1}{d_i}
\sum_{j=1}^{d_i}
\left(
\frac{\hat{c}_{ij}-c_{ij}}{c_{ij}}
\right)^2,
\]
where each error is normalized by the ground-truth coefficient itself. As a numerical safeguard, a ground-truth coefficient with \(|c_{ij}|<10^{-10}\) contributes zero error when its prediction is also below \(10^{-10}\); otherwise, the sample is treated as non-evaluable and excluded from \(\mathcal{I}_{c}\). This metric is not computed for predictions with unmatched symbolic type or unsupported coefficient extraction. Unless stated otherwise, \(\hat{\mathbf{c}}_i\) denotes the coefficients used for reconstruction after coefficient refinement. Experiments that evaluate the raw decoder output before BFGS identify that setting explicitly.
Because \(|\mathcal{I}_{c}|\) can differ across models and evaluation conditions, MRCSE comparisons are accompanied by their evaluable sample counts.

\paragraph{Relative \(L^2\) error.}
To measure trajectory-level fidelity, we solve or reconstruct the PDE implied by the predicted expression and compare the resulting trajectory \(\hat{\mathbf{u}}_i\) with the ground-truth trajectory \(\mathbf{u}_i\). For the reconstructable set \(\mathcal{I}_{u}\), we report
\[
\mathrm{Rel}\, L^2
=
\frac{1}{|\mathcal{I}_{u}|}
\sum_{i\in\mathcal{I}_{u}}
\frac{\|\hat{\mathbf{u}}_i-\mathbf{u}_i\|_2}{\|\mathbf{u}_i\|_2}.
\]
This metric evaluates agreement in solution space over the observed time window. In the corruption experiment, the prediction is generated from the corrupted augmented input, but the decoded PDE is reconstructed from the corresponding clean initial condition and compared with the noise-free trajectory. Relative \(L^2\) therefore measures recovery of the underlying clean dynamics rather than agreement with the injected perturbation. It is complementary to TRR: low relative \(L^2\) error does not by itself imply correct symbolic discovery.

\paragraph{Uncertainty intervals.}
Multiple trajectories can share the same completed symbolic equation because each coefficient setting is solved under several initial conditions. We therefore treat the coefficient-instantiated equation, rather than an individual trajectory, as the independent resampling unit. For the clean in-domain and noise experiments, we apply a percentile cluster bootstrap with 10,000 replicates to 256 coefficient settings, each represented by 20 initial conditions. For strict parameter OOD, we resample 640 coefficient settings, each represented by eight initial conditions. Missing decoder outputs are counted as failures in TRR; relative \(L^2\) error and MRCSE retain the evaluable supports defined above. The resulting 95\% intervals quantify variation across the finite test equations for a fixed pretrained checkpoint. They do not measure variation across training seeds or checkpoint selection.

\subsection{Few-shot fine-tuning}
\label{sec:few_shot_finetuning}

Few-shot fine-tuning is used to test whether the pretrained decoder can be adapted to new PDE families that were not present as complete symbolic forms during pretraining. The current manuscript evaluates four held-out composite targets:
\[
\begin{aligned}
&u_t = -\alpha u_x - k u\,u_x + \nu u_{xx}, &&\text{(advection--Burgers)}\\
&u_t = -\alpha u_x - k u^2\,u_x + \nu u_{xx}, &&\text{(advection--cubic-Burgers)}\\
&u_t = -\alpha u_x - k\cos(u)\,u_x + \nu u_{xx}, &&\text{(advection--sine)}\\
&u_t = -\alpha u\,u_x - k u^2\,u_x + \nu u_{xx}. &&\text{(Burgers--cubic-Burgers)}
\end{aligned}
\]
Each target combines transport, nonlinear flux, and diffusion terms that appear individually in the pretraining corpus, but not as the full composite symbolic form.

For the few-shot adaptation protocol, all four target-family runs start from the same separately trained FoundPDE checkpoint, obtained after 100 epochs of pretraining on 512,000 solution--expression samples from the original PDE corpus. This transfer checkpoint shares the training corpus and schedule of the principal model used for the preceding experiments but takes only the solution field as input, without Fourier or derivative augmentation. We define fine-tuning epoch 0 as the model state immediately after loading this transfer checkpoint and before any target-family optimizer update. A separate copy of that state is then fine-tuned on each target family and evaluated on held-out examples from that family. Advection--Burgers uses 1000 target-family training samples, while each of the other three families uses 500 training samples. Evaluations contain 100 expected test samples per family, and outputs that cannot be parsed are counted as incorrect. For the advection--Burgers run only, we also measure retention on the 20 original pretraining families after fine-tuning. Retention is reported as the unweighted macro-average of the 20 family-level TRRs, using each family's fixed expected sample count and treating missing or unparsable outputs as incorrect. This retention test is important because a successful adaptation method should improve new-family recovery without erasing the pretrained discovery capability. The composite-family equation forms, coefficient ranges, initial-condition generator, source-checkpoint identity, and complete comparison protocol are documented in Appendix~\ref{app:custom_composite_setup}.

As context, we compare against one model trained from scratch jointly on the same four composite PDE families. This run uses 25 coefficient settings and 20 initial conditions per family for training, and 5 coefficient settings and 20 initial conditions per family for testing. For advection--cubic-Burgers, advection--sine, and Burgers--cubic-Burgers, the single-family fine-tuning files are exact family-wise subsets of this mixed scratch dataset. Both protocols use 10 optimizer updates per epoch, but they differ in batch composition and training objective: fine-tuning adapts one checkpoint to one family at a time, whereas scratch training learns all four families jointly. The comparison therefore tests whether the reported adaptation protocol reaches useful recovery over a shorter training horizon; it is not an initialization-only ablation with otherwise identical optimization conditions.

\subsection{Unseen PDE prediction with hierarchical discovery}
\label{sec:hierarchical_discovery}

A general and robust foundation model for PDE discovery requires a large corpus of diverse training data. Building such a corpus is computationally expensive because every coefficient setting and initial condition requires a new numerical simulation or experimental observation. Moreover, combining differential terms, nonlinearities, derivative orders, and coefficients can produce millions of candidate PDE types, making exhaustive data generation and pretraining impractical. Consequently, even a broad FoundPDE corpus can cover only a small subset of the possible symbolic structures, and direct decoding may fail when a test equation contains a combination of terms that was not included during training.

The compositional structure of PDEs nevertheless offers an opportunity to use the pretrained foundation model more fully. An out-of-distribution PDE may be new as a complete symbolic expression while still being assembled from operators that FoundPDE learned individually. We therefore propose the \emph{hierarchical discovery algorithm} to extend a pretrained FoundPDE model to a new class of unseen PDE types: additive two-operator equations whose constituent operators are individually represented in pretraining and whose dynamics admit a dominant--subordinate separation. The algorithm identifies the dominant familiar operator, removes its estimated contribution from the observed evolution, applies FoundPDE again to the residual dynamics to recover the subordinate operator, and finally combines and jointly calibrates the two recovered terms.

\begin{figure}[htbp]
    \centering
    \includegraphics[width=\linewidth]{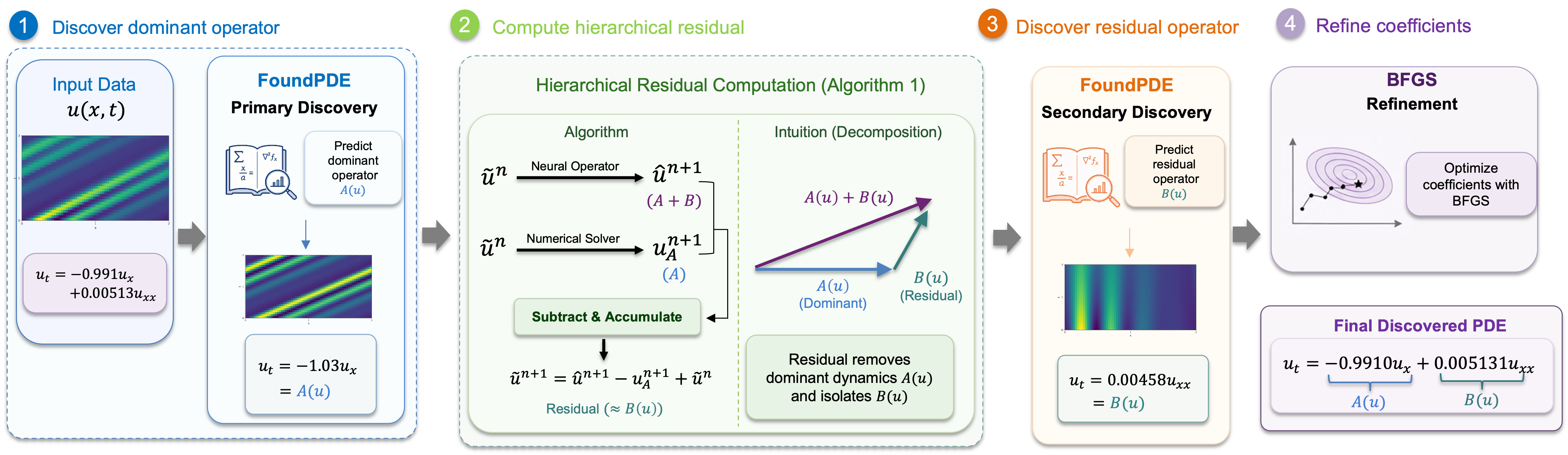}
            
        \caption{\textbf{The hierarchical discovery algorithm sequentially recovers a composite PDE from familiar operators.} The first FoundPDE pass identifies the dominant operator \(u_t=A(u)\) from a trajectory of the composite equation \(u_t=A(u)+B(u)\). The residual-computation panel depicts one update at time step \(n\): the neural operator supplies \(\hat{u}^{\,n+1}\), the dominant-equation solve supplies \(u_A^{\,n+1}\), and their difference updates \(\tilde{u}^{\,n+1}\). Repeating this calculation over all time steps constructs the residual dataset \(\tilde{\mathcal{D}}\), as formalized in Algorithm~\ref{alg:hierarchical_discovery}. A second FoundPDE pass identifies \(B(u)\), after which the two operators are combined and calibrated with BFGS.}
    \label{fig:hierarchical_discovery}
\end{figure}

To illustrate the idea, we consider a model pretrained on the individual PDE forms \(u_t=A(u)\) and \(u_t=B(u)\), but not on their composite \(u_t=A(u)+B(u)\). The design uses the model's tendency to identify the dynamically dominant component in the first pass. To explain the method, we consider one observed trajectory and index only its time steps by \(n\). After FoundPDE recovers \(A(u)\), a neural operator is trained and predicts the dynamics update \(\hat{u}^{\,n+1}\) under \(A(u)+B(u)\), while a numerical solve of the recovered dominant equation computes \(u_A^{\,n+1}\) under \(A(u)\) alone. Their difference isolates the incremental contribution of \(B(u)\); adding this difference to the current residual state \(\tilde{u}^{\,n}\) produces the next residual state \(\tilde{u}^{\,n+1}\). FoundPDE is then applied to the resulting residual trajectory to recover the subordinate operator, after which the two operators are combined and their coefficients are jointly refined with BFGS. The complete sequence is formalized in Algorithm~\ref{alg:hierarchical_discovery}, while Fig.~\ref{fig:hierarchical_discovery} provides a schematic overview.

\begin{algorithm}[htbp]
\caption{\textbf{Hierarchical discovery for sequential operator recovery.} Assume one solution dataset \(\mathcal{D}=\{u^n\}_{n=0}^{N_t-1}\) from the composite equation \(u_t=A(u)+B(u)\).}
\label{alg:hierarchical_discovery}
\begin{algorithmic}[1]
    \State Train a neural operator \(\mathcal{N}\) on \(\mathcal{D}\) to model the next-timestep evolution of \(u_t=A(u)+B(u)\).
    \State Apply FoundPDE to \(\mathcal{D}\) and infer the dominant operator \(u_t=A(u)\).
    \State Initialize the residual trajectory with \(\tilde{u}^{\,0}=u^0\).
    \For{\(n=0,\ldots,N_t-2\)}
        \State Compute the full-dynamics update \(\hat{u}^{\,n+1}=\mathcal{N}(\tilde{u}^{\,n})\).
        \State Advance the dominant equation \(u_t=A(u)\) from \(\tilde{u}^{\,n}\) by one step to obtain \(u_A^{\,n+1}\).
        \State Set \(\tilde{u}^{\,n+1}=\hat{u}^{\,n+1}-u_A^{\,n+1}+\tilde{u}^{\,n}\).
    \EndFor
    \State Store \(\tilde{\mathcal{D}}=\{\tilde{u}^{\,n}\}_{n=0}^{N_t-1}\).
    \State Apply FoundPDE to \(\tilde{\mathcal{D}}\) and infer the subordinate operator \(u_t = B(u)\).
    \State Combine the recovered operators as \(u_t=A(u)+B(u)\).
    \State Jointly refine all coefficients in the combined expression with BFGS.
\end{algorithmic}
\end{algorithm}

The residual construction in Algorithm~\ref{alg:hierarchical_discovery} follows an operator-splitting interpretation. At time step \(n\), the neural-operator prediction \(\hat{u}^{\,n+1}=\mathcal{N}(\tilde{u}^{\,n})\) approximates the combined evolution under \(A(u)+B(u)\), whereas \(u_A^{\,n+1}\) is obtained by advancing the recovered dominant equation and therefore represents the evolution under \(A(u)\) alone. The update
$
\tilde{u}^{\,n+1}=\hat{u}^{\,n+1}-u_A^{\,n+1}+\tilde{u}^{\,n}
$
removes the dominant one-step contribution and accumulates the remaining increment. Iterating over \(n\) produces the residual dataset \(\tilde{\mathcal{D}}=\{\tilde{u}^{\,n}\}_{n=0}^{N_t-1}\), which is supplied to the second FoundPDE pass to recover \(B(u)\).

The surrogate \(\mathcal{N}\) is essential because it provides the full-dynamics reference against which the dominant contribution is subtracted. We instantiate it as a per-target Fourier Neural Operator (FNO)~\cite{li2021fourierneuraloperatorparametric}. Errors in \(\hat{u}^{\,n+1}\) propagate directly into the residual sequence, so the accuracy of the second-stage discovery depends on both the quality of the recovered dominant operator and the one-step surrogate. If the single observed trajectory contains many time steps, it provides many consecutive input--output pairs and helps train the neural operator. In practice, however, one trajectory may not provide enough diversity for an accurate surrogate. Additional trajectories generated or observed under the same governing equation, for example from different initial conditions, can therefore be included when training \(\mathcal{N}\). The FNO training data and discretization, together with the procedure for aggregating and resolving ambiguous first-stage predictions, are detailed in Appendix~\ref{app:hierarchical_discovery_details}.

We test the hierarchical discovery algorithm using a separate model pretrained only on advection, inviscid Burgers, inviscid cubic Burgers, and diffusion, and evaluate three unseen two-term composites: advection--diffusion, viscous Burgers, and viscous cubic Burgers. Although the method above is presented with one trajectory for clarity, each experiment uses \(M = 20\) trajectories with independently sampled initial conditions to provide sufficient data for neural-operator training and to evaluate dominant-operator recovery, diffusion-residual recovery, and BFGS coefficient calibration. Results are reported in Sec.~\ref{experiment:hierarchical_discovery}, and the concrete trajectory count and full experimental protocol are provided in Appendix~\ref{app:hierarchical_discovery_details}.

\section{Results}
\label{sec:experiments}

This section evaluates FoundPDE in a sequence of symbolic-discovery settings. We first consider in-domain symbolic recovery and feature augmentation, followed by per-family behavior and representative qualitative examples. We then test robustness to corruption of the augmented model input and compare with SINDy and DISCOVER under sparse-data protocols. Finally, we evaluate parameter-OOD generalization within known symbolic families and study extension beyond the pretraining vocabulary through few-shot fine-tuning and hierarchical discovery.

\subsection{Dataset and experimental setup}

In our experiments, we construct a dataset of symbolic PDE expressions following the same formulation as in the PROSE-PDE work~\cite{sun2025foundationmodelpartialdifferential}. Specifically, we first generate 25.6K symbolic forms representing 20 distinct PDE types. For each generated PDE, we solve the equation using numerical solvers under 20 different initial conditions, resulting in a total of 512K solution samples. Detailed descriptions of the PDE generation process, initial condition design, and numerical solution methods are provided in Appendix~\ref{app:dataset_specification} and Appendix~\ref{app:initial_conditions}.
The FoundPDE model is trained on this dataset using a transformer-based sequence-to-sequence architecture. Architecture, optimizer, decoding, feature-augmentation, and hardware details are reported in Appendix~\ref{app:model_training_details}.

The symbolic forms used for training and in-domain testing are produced by the same three-stage generator: each PDE type defines a skeleton, coefficients are sampled to instantiate a complete symbolic equation, and an initial condition is sampled before numerical solution. Training and test examples are sampled independently. They therefore share the same generative distribution but, because both coefficients and initial conditions are randomized, do not intentionally reuse trajectories or completed symbolic instances.
For the in-domain corpus, positive-valued coefficients are sampled within the relative support \([0.1,1.9]\) around their family-specific reference values, while the signed advection coefficient is sampled over \([-1.9,1.9]\). Appendix Table~\ref{tab:appendix_pde_families} summarizes the family-specific reference values and in-domain supports.

We train FoundPDE in four modes described in Sec.~\ref{sec:framework}: barebone, Fourier-augmented, spatial-derivative-augmented, and Fourier-and-derivative-augmented. We evaluate the model under the following conditions:
\begin{enumerate}
    \item \textbf{In-domain testing (Sec.~\ref{sec:in_domain}):} Evaluating PDEs with parameter values within the training range.
    \item \textbf{Robustness testing (Sec.~\ref{sec:noise_robustness}):} Evaluating whether symbolic recovery remains meaningful under Gaussian corruption of the augmented model input.
    \item \textbf{Baseline comparison (Sec.~\ref{sec:baseline_comparison}):} Comparing FoundPDE with existing symbolic PDE discovery methods under sparse-grid protocols.
    \item \textbf{Parameter-OOD generalization (Sec.~\ref{sec:ood_parameter}):} Testing known symbolic PDE families with coefficient ranges outside the nominal in-domain support.
    \item \textbf{Few-shot fine-tuning (Sec.~\ref{sec:few_shot_results}) and hierarchical discovery (Sec.~\ref{experiment:hierarchical_discovery}):} Evaluating targeted adaptation and selected compositional recovery for PDE structures outside the direct pretraining support.
\end{enumerate}

\subsection{In-domain performance}
\label{sec:in_domain}

\begin{figure}[htbp]
\centering
\includegraphics[width=\linewidth]{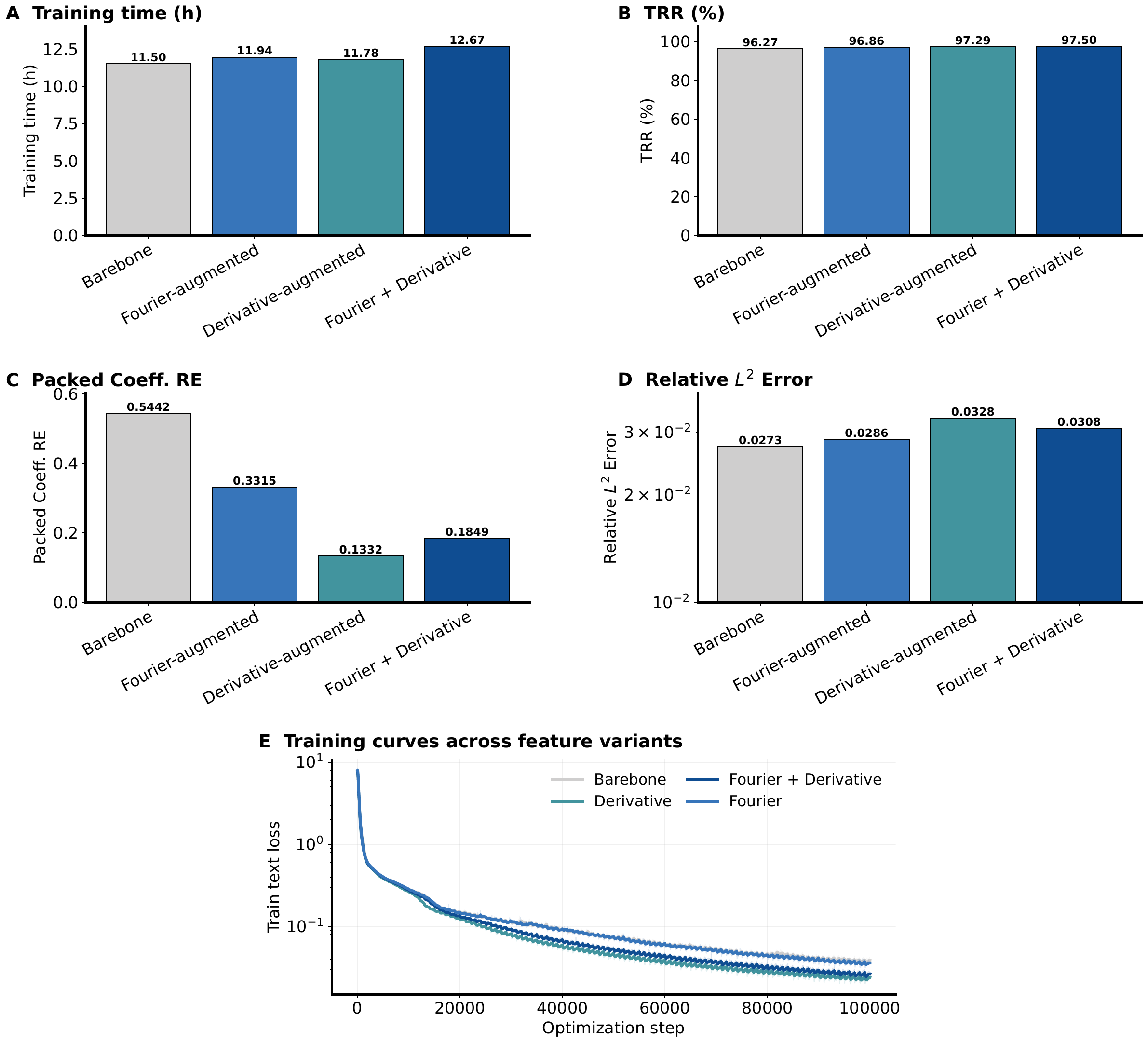}
\caption{\textbf{Effects of feature augmentation across symbolic and numerical metrics.} All four variants are trained on the same 512,000 samples for 100 epochs. (\textbf{A}) Projected training time for the four model variants, from the measured time per optimization step (Appendix~\ref{app:model_training_details}). (\textbf{B}) The augmented models improve type recognition rate (TRR) relative to the barebone baseline. The Fourier-and-derivative model gives the highest structural accuracy. (\textbf{C}) The derivative model gives the lowest mean relative coefficient squared error (MRCSE). The evaluable MRCSE counts are 4,929, 4,959, 4,980, and 4,992 for the barebone, Fourier, derivative, and Fourier-and-derivative models, respectively. (\textbf{D}) The barebone model gives the lowest mean relative $L^2$ error; the evaluable count is 5,115 for every variant. (\textbf{E}) The training text-loss curves indicate that the final training loss is not by itself a reliable proxy for downstream symbolic recovery.}
\label{fig:in_domain_feature_training_combo}
\end{figure}

We first evaluate whether FoundPDE can recover the symbolic structure and coefficients of PDEs whose types and parameter ranges are represented in the training distribution. Fig.~\ref{fig:in_domain_feature_training_combo} summarizes the four feature variants on clean in-domain data. All four variants are trained on the same 512,000 samples with the same optimization schedule. Figs.~\ref{fig:in_domain_feature_training_combo}A--D show that the augmented inputs improve symbolic recovery under the in-domain setting, although the gains are small at this training budget. Missing decoder outputs are counted as failures for every feature variant. The Fourier-and-derivative model reaches the highest TRR, \(97.50\%\), compared with \(97.29\%\) for the derivative model, \(96.86\%\) for the Fourier model, and \(96.27\%\) for the barebone model. Its clustered 95\% bootstrap interval is \([96.54\%,98.30\%]\). Because all variants are evaluated on the same test samples, we also bootstrap the paired TRR differences with the same coefficient-setting clusters. The Fourier-and-derivative model exceeds the barebone model by \(1.23\) percentage points (95\% interval \([0.64,1.88]\)) and the Fourier model by \(0.64\) points (\([0.25,1.05]\)), whereas its \(0.21\)-point margin over the derivative model is within test-set variation (\([-0.27,0.70]\)). The first-order spatial derivative therefore accounts for most of the structural gain. The augmented inputs add little training cost at this scale: the projected training time increases from \(11.50\) hours for the barebone model to \(12.67\) hours for the Fourier-and-derivative model. The numerical metrics follow a different ordering: the derivative-augmented model gives the lowest MRCSE, \(0.133\), while the barebone model gives the lowest mean relative \(L^2\) error, \(0.0273\). For the selected model, the mean relative \(L^2\) error is \(0.0308\), with a 95\% interval of \([0.0241,0.0384]\). These results indicate that symbolic structure recovery, MRCSE, and relative \(L^2\) error measure distinct aspects of discovery quality. Appendix Table~\ref{tab:cluster_bootstrap_intervals} reports the complete uncertainty analysis. Fig.~\ref{fig:in_domain_feature_training_combo}E provides the corresponding optimization curves. The derivative-augmented model reaches the lowest final training text loss, whereas the Fourier-and-derivative model gives the highest TRR. Thus, in this setting, the training objective is correlated with but not identical to the final symbolic-discovery metric.

\begin{figure}[htbp]
\centering
\includegraphics[width=\linewidth]{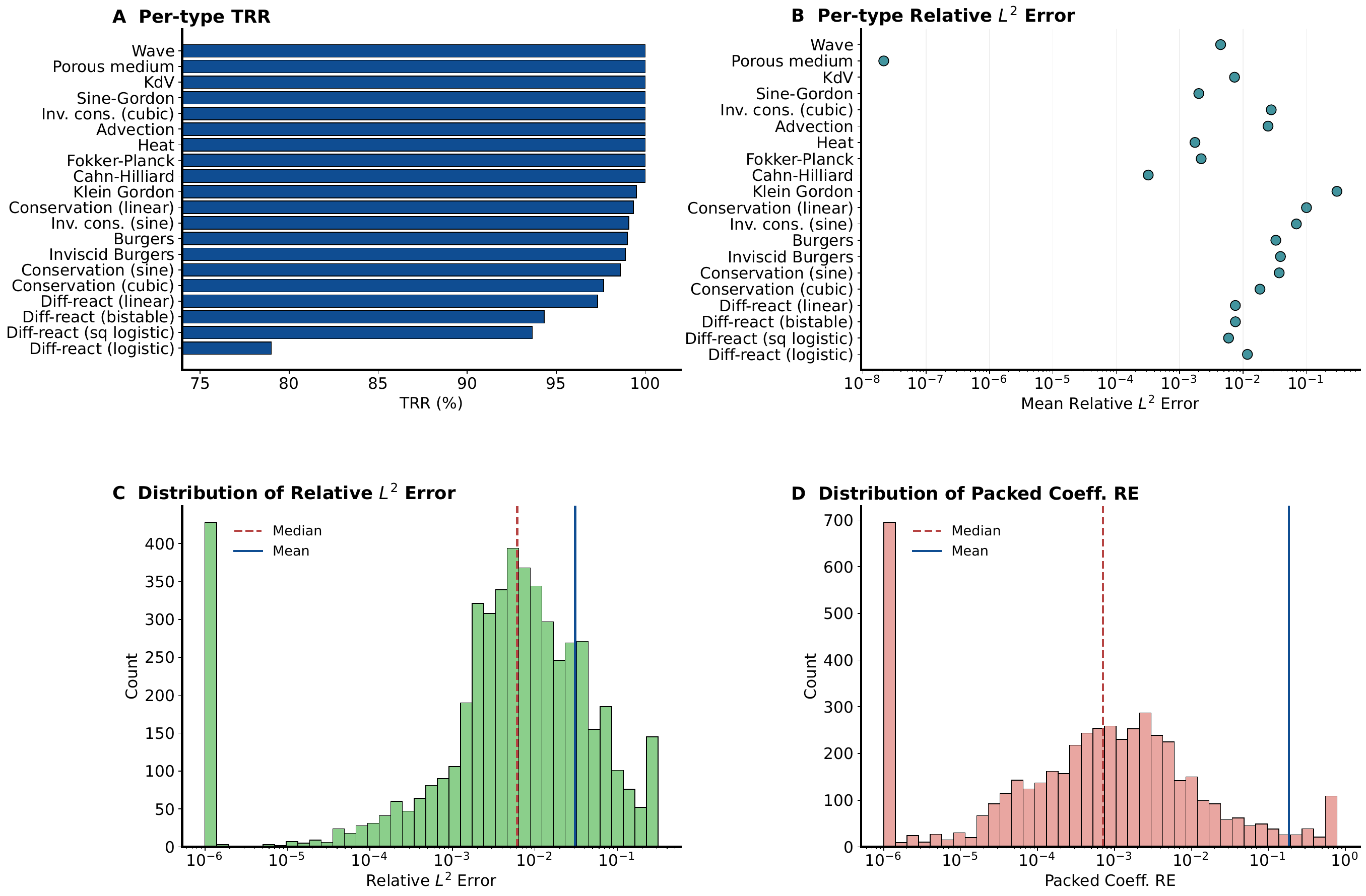}
\caption{\textbf{In-domain recovery varies across PDE families and evaluation criteria.} (\textbf{A}) Per-type TRR. Nine families, including heat, advection, porous medium, KdV, wave, Sine--Gordon, and Fokker--Planck, are recognized for every test sample, whereas the diffusion-reaction families have the lowest TRR. (\textbf{B}) Mean relative $L^2$ error by family. The family ordering differs from that in panel A, showing that structural recognition and relative $L^2$ error need not rank PDE families identically. (\textbf{C}) Global relative $L^2$ error distribution over 5,115 reconstructable predictions. (\textbf{D}) Global MRCSE distribution over 4,992 type-matched predictions with extractable coefficients; its long tail shows that a subset of evaluable predictions remains difficult to calibrate.}
\label{fig:in_domain_final_report}
\end{figure}

After comparing feature variants, we examine the final selected model by PDE family. Fig.~\ref{fig:in_domain_final_report} reports per-family TRR and relative \(L^2\) error, together with the global distributions of relative \(L^2\) error and MRCSE.
Fig.~\ref{fig:in_domain_final_report}A shows that nine families, including heat, advection, wave, Sine--Gordon, porous medium, KdV, and Fokker--Planck, are recognized for every test sample. Diffusion-reaction equations have the lowest structural recovery rates, with the logistic reaction family the most difficult case at \(79.0\%\); every other family exceeds \(93\%\). Comparing Figs.~\ref{fig:in_domain_final_report}A and B shows that structural and trajectory metrics do not induce the same family ordering. Klein--Gordon reaches \(99.5\%\) TRR but has mean relative \(L^2\) error \(0.304\), whereas the logistic and bistable diffusion-reaction families have lower TRR, \(79.0\%\) and \(94.3\%\), but mean relative \(L^2\) errors below \(0.012\). Figs.~\ref{fig:in_domain_final_report}C and D further show that the global numerical-error distributions are right-skewed. We therefore report symbolic, coefficient, and trajectory metrics separately rather than treating any one metric as a proxy for the others.

\begin{figure}[htbp]
\centering
\begin{subfigure}[t]{0.98\textwidth}
    \centering
    \includegraphics[width=\linewidth]{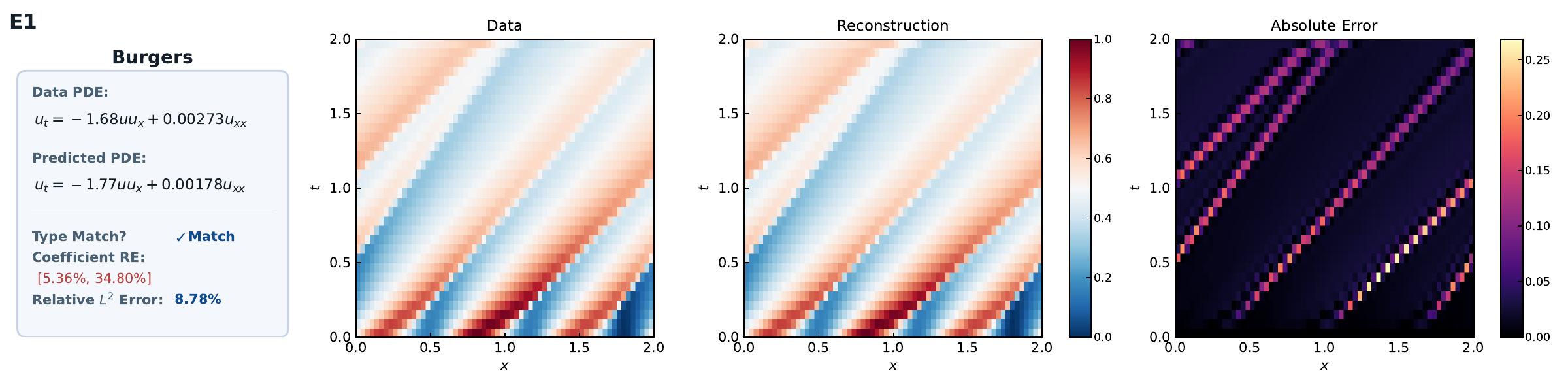}
\end{subfigure}

\vspace{-1.1em}

\begin{subfigure}[t]{0.98\textwidth}
    \centering
    \includegraphics[width=\linewidth]{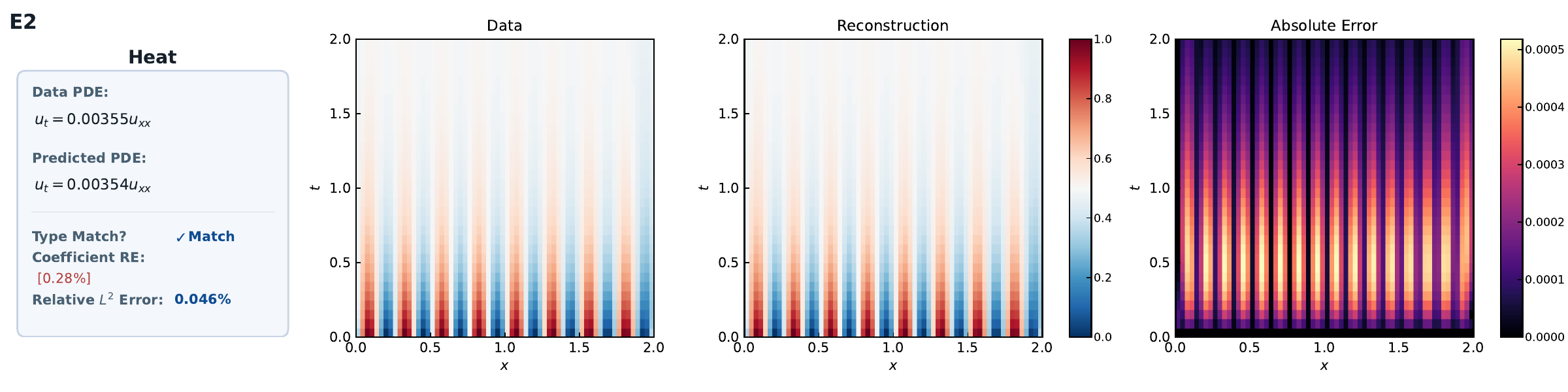}
\end{subfigure}

\vspace{-1.1em}

\begin{subfigure}[t]{0.98\textwidth}
    \centering
    \includegraphics[width=\linewidth]{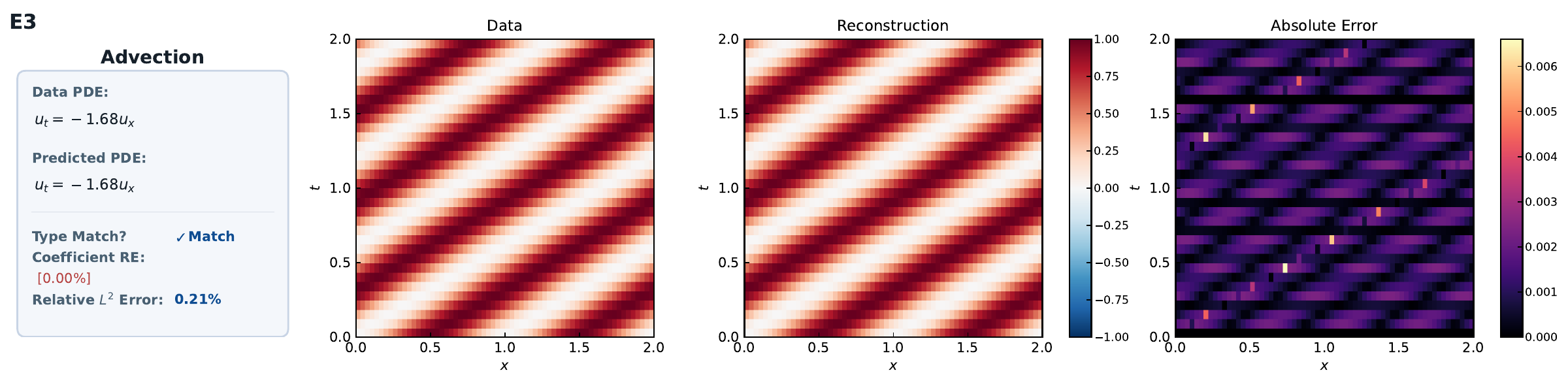}
\end{subfigure}

\vspace{-1.1em}

\begin{subfigure}[t]{0.98\textwidth}
    \centering
    \includegraphics[width=\linewidth]{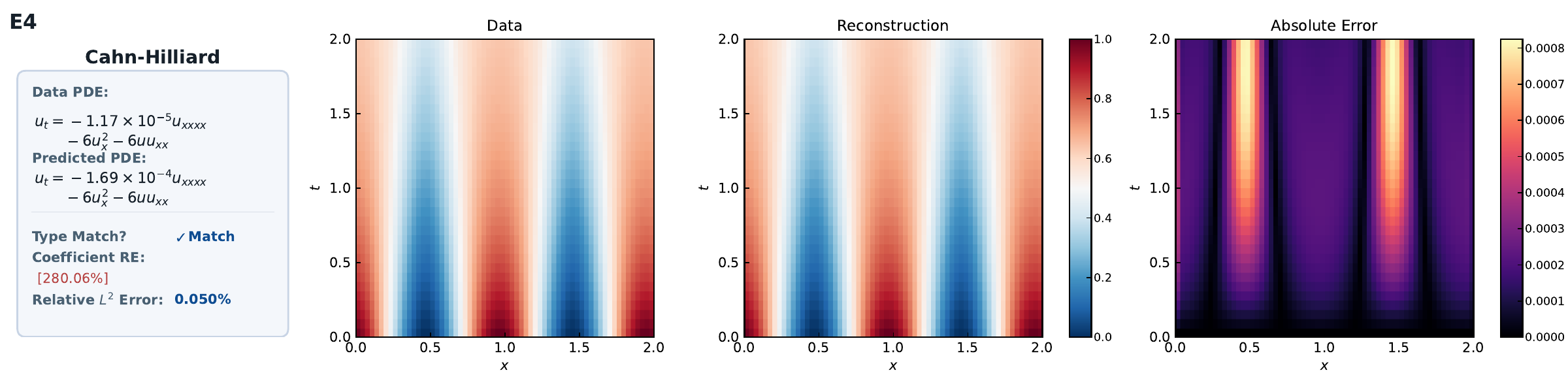}
\end{subfigure}

\vspace{-1.1em}

\begin{subfigure}[t]{0.98\textwidth}
    \centering
    \includegraphics[width=\linewidth]{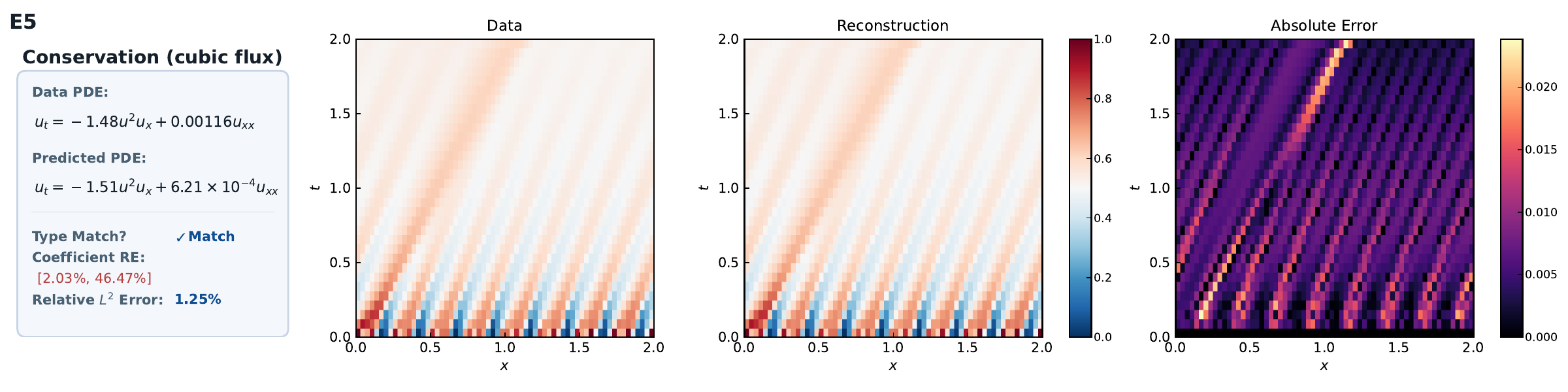}
\end{subfigure}
\caption{\textbf{Representative in-domain examples E1--E5 show accurate symbolic recovery on transport, diffusion, and higher-order families.} This group covers Burgers, heat, advection, Cahn--Hilliard, and conservation with cubic flux. All five predictions have the correct symbolic type, and the reconstructed fields closely match the ground truth. E4 recovers the Cahn--Hilliard structure with a visibly different fourth-order coefficient while retaining a small relative \(L^2\) error.}
\label{fig:triptych_pair_1}
\end{figure}

\begin{figure}[htbp]
\centering
\begin{subfigure}[t]{0.98\textwidth}
    \centering
    \includegraphics[width=\linewidth]{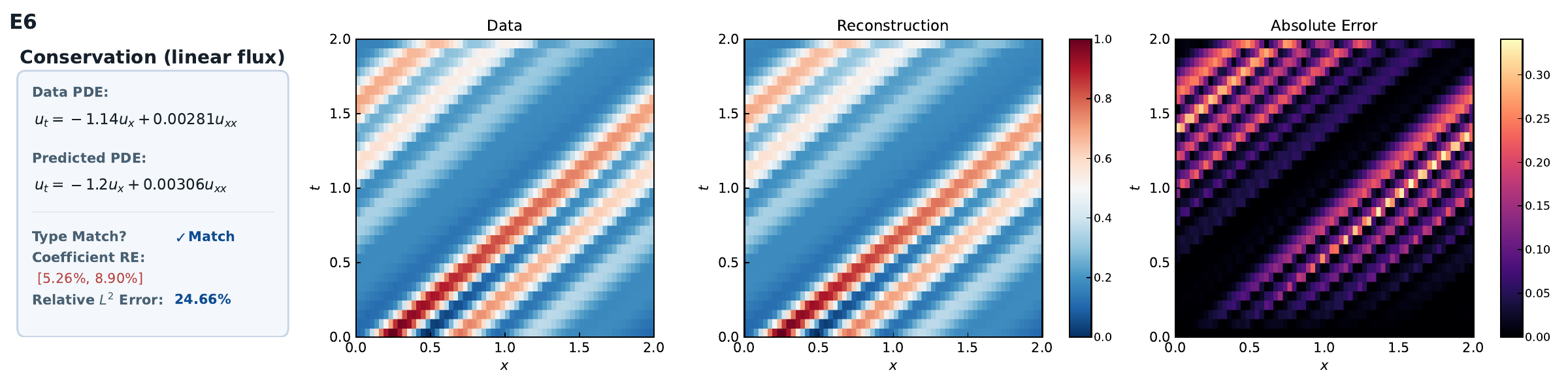}
\end{subfigure}

\vspace{-1.1em}

\begin{subfigure}[t]{0.98\textwidth}
    \centering
    \includegraphics[width=\linewidth]{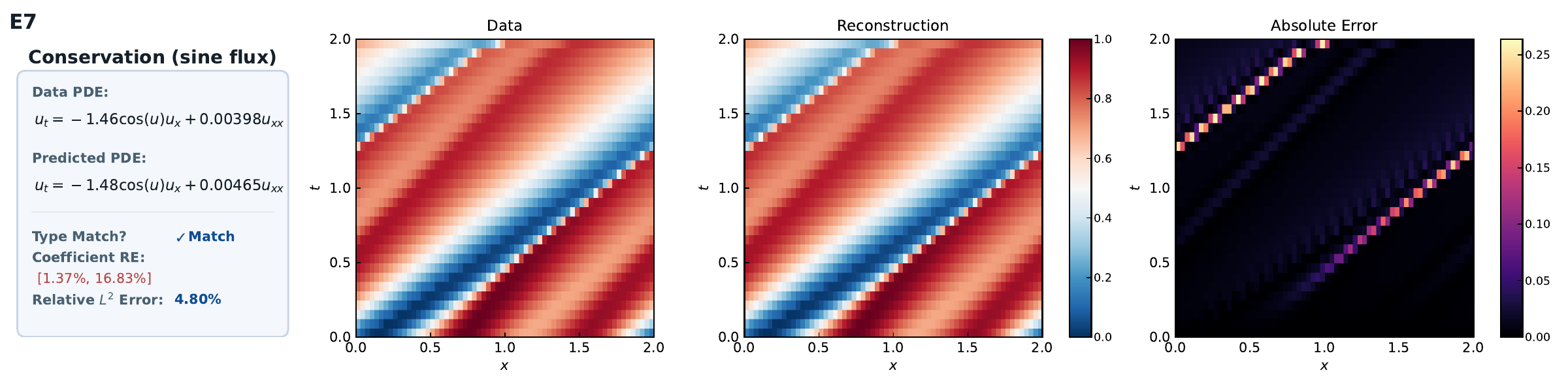}
\end{subfigure}

\vspace{-1.1em}

\begin{subfigure}[t]{0.98\textwidth}
    \centering
    \includegraphics[width=\linewidth]{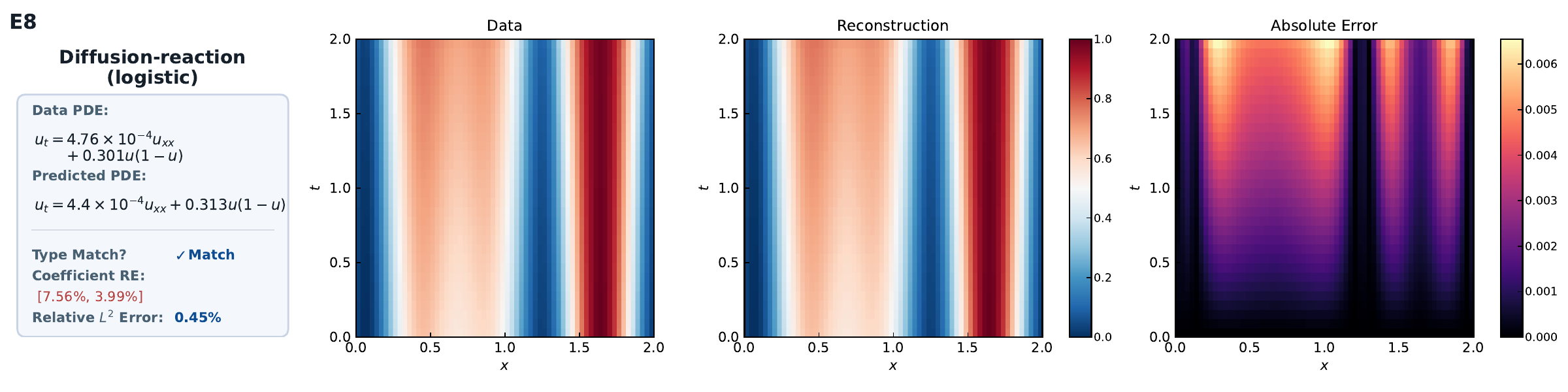}
\end{subfigure}

\vspace{-1.1em}

\begin{subfigure}[t]{0.98\textwidth}
    \centering
    \includegraphics[width=\linewidth]{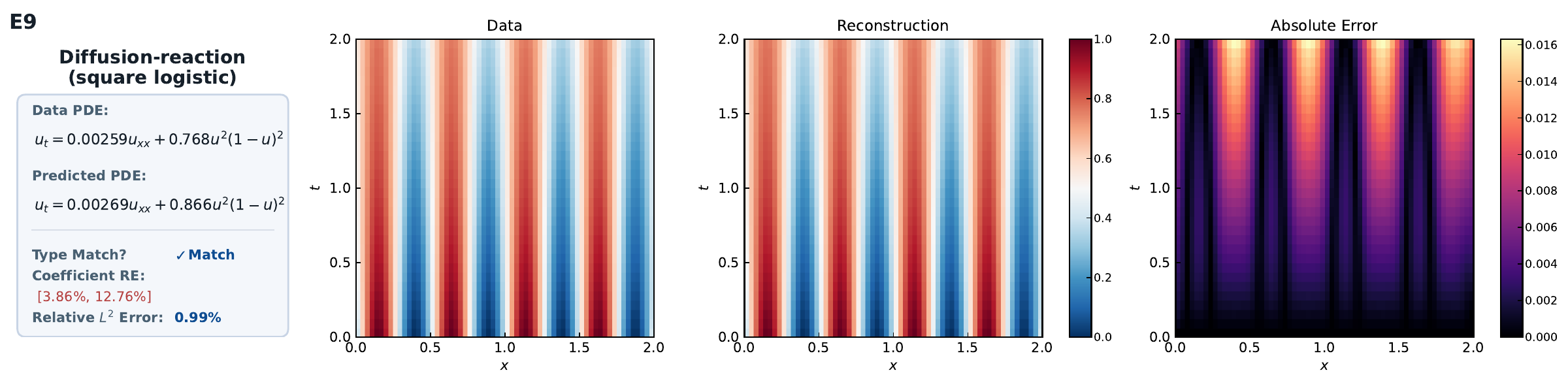}
\end{subfigure}

\vspace{-1.1em}

\begin{subfigure}[t]{0.98\textwidth}
    \centering
    \includegraphics[width=\linewidth]{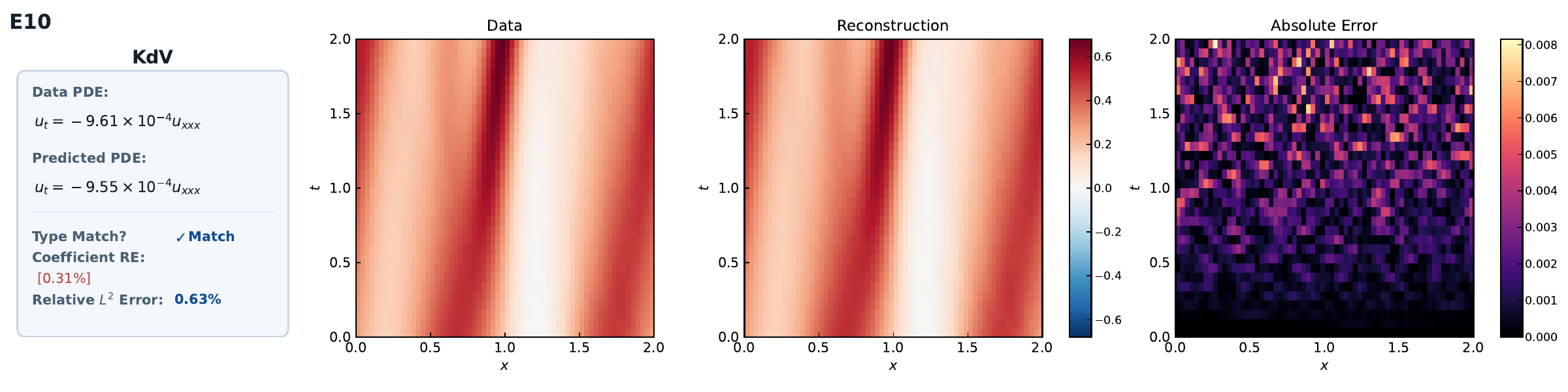}
\end{subfigure}
\caption{\textbf{Representative in-domain examples E6--E10 separate structural recovery from numerical calibration.} This set includes linear-flux conservation, sine-flux conservation, logistic reaction-diffusion, square-logistic reaction-diffusion, and KdV. All five predictions have the correct symbolic type. E7--E10 agree well with the ground truth in field space, whereas small errors in the transport and diffusion coefficients of E6 produce a visible phase shift.}
\label{fig:triptych_pair_2}
\end{figure}

\begin{figure}[htbp]
\centering
\begin{subfigure}[t]{0.98\textwidth}
    \centering
    \includegraphics[width=\linewidth]{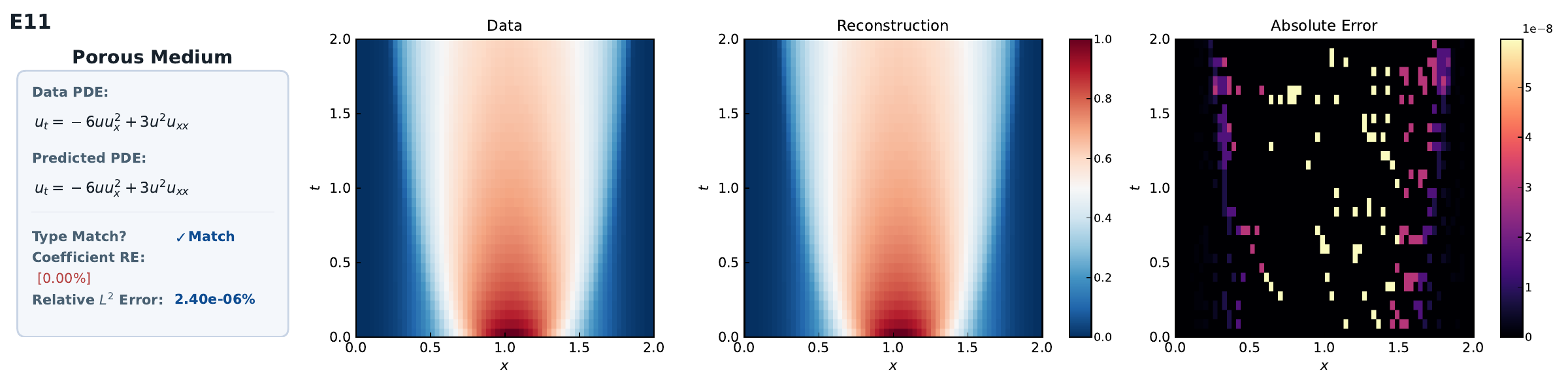}
\end{subfigure}

\vspace{-1.1em}

\begin{subfigure}[t]{0.98\textwidth}
    \centering
    \includegraphics[width=\linewidth]{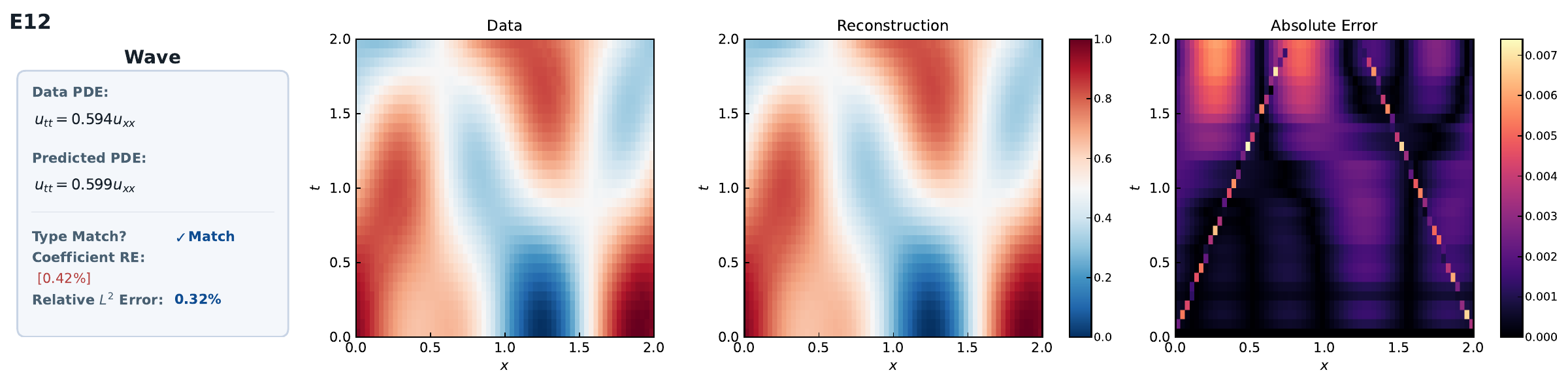}
\end{subfigure}

\vspace{-1.1em}

\begin{subfigure}[t]{0.98\textwidth}
    \centering
    \includegraphics[width=\linewidth]{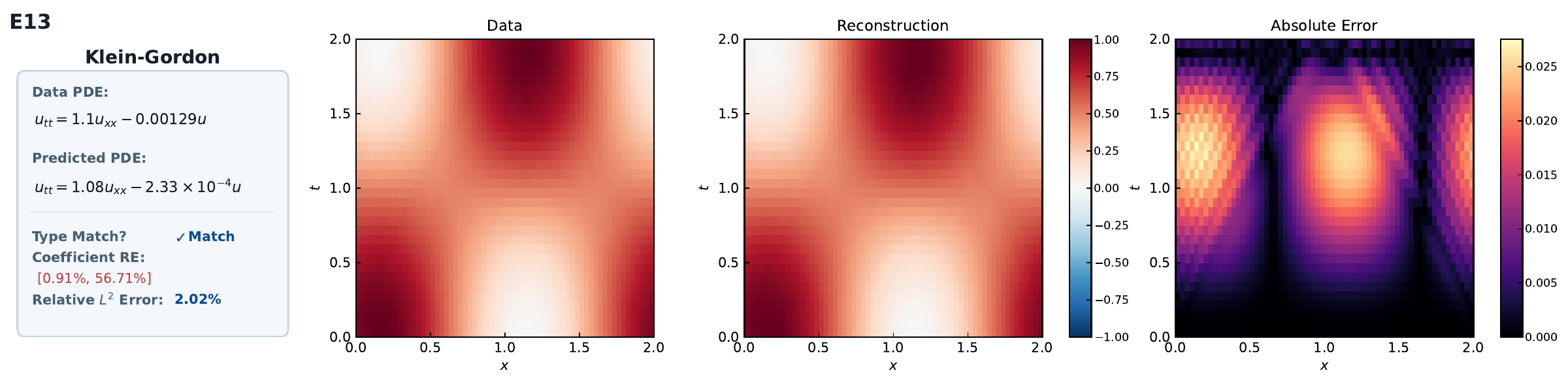}
\end{subfigure}

\vspace{-1.1em}

\begin{subfigure}[t]{0.98\textwidth}
    \centering
    \includegraphics[width=\linewidth]{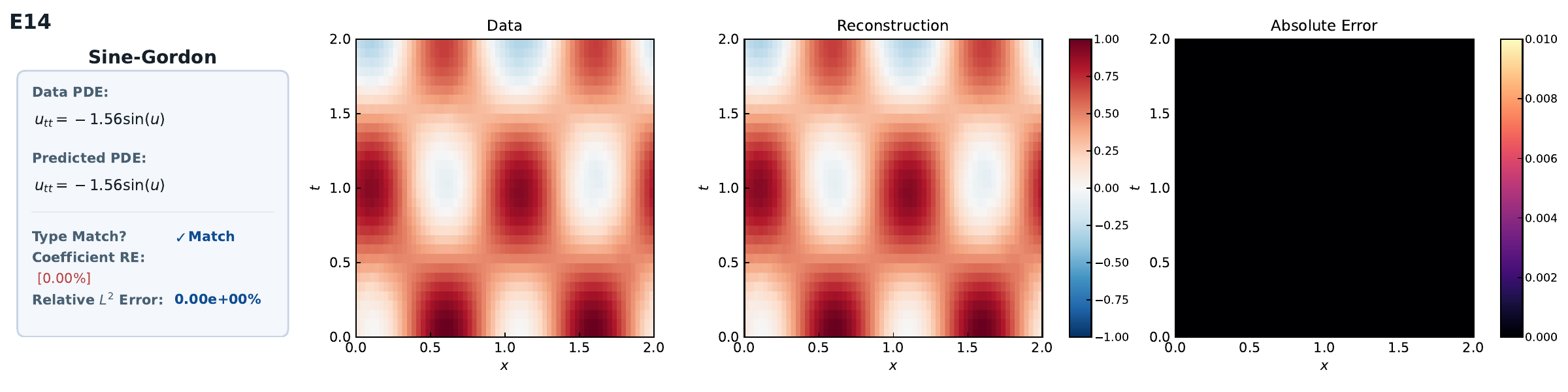}
\end{subfigure}

\vspace{-1.1em}

\begin{subfigure}[t]{0.98\textwidth}
    \centering
    \includegraphics[width=\linewidth]{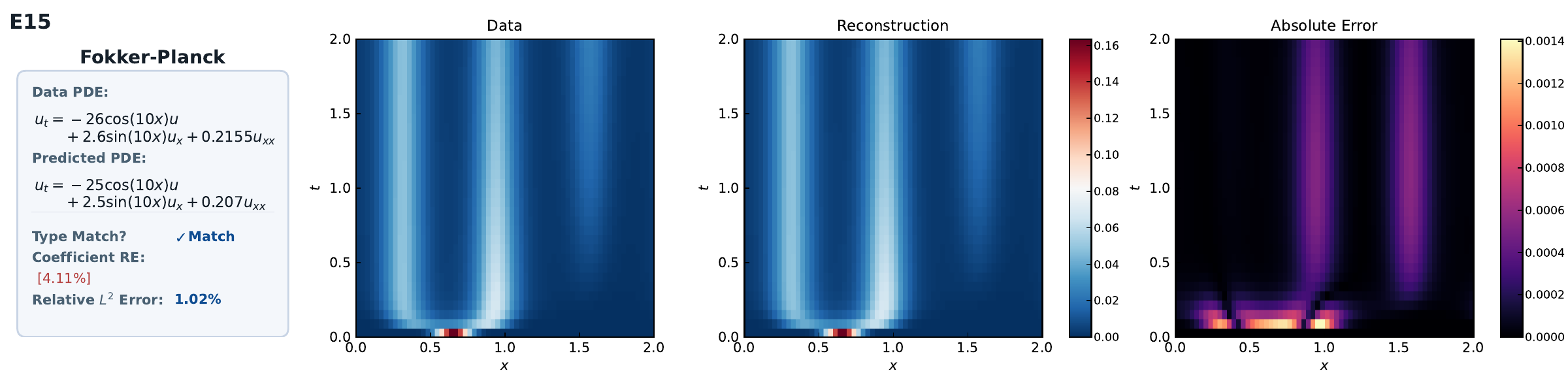}
\end{subfigure}
\caption{\textbf{Representative in-domain examples E11--E15 show that FoundPDE handles porous-medium, hyperbolic, and driven-diffusion equations well.} This group contains porous medium, wave, Klein--Gordon, Sine--Gordon, and Fokker--Planck examples. The porous-medium, wave, Sine--Gordon, and Fokker--Planck cases show close visual agreement between predicted and ground-truth fields, while Klein--Gordon illustrates a regime where the symbolic type is correct but coefficient sensitivity is still visible.}
\label{fig:triptych_pair_3}
\end{figure}

To complement the aggregate metrics, we then provide an example-by-example visual comparison in Figs.~\ref{fig:triptych_pair_1}--\ref{fig:triptych_pair_3}. Each triptych contains: (i) the ground-truth and predicted symbolic PDEs with their numerical coefficients, (ii) the ground-truth and predicted solution fields over the $(x,t)$ domain, and (iii) the absolute error map. These examples show representative recoveries as well as cases where symbolic and relative \(L^2\) metrics differ.
All 15 displayed predictions have the correct symbolic type, and correct symbolic predictions generally lead to close agreement between predicted and ground-truth fields. In Examples E2, E3, E11, and E14, the absolute error remains small over most of the space-time domain. The examples also show that coefficient errors and relative \(L^2\) errors need not agree. In E4, the decoded fourth-order Cahn--Hilliard coefficient is \(1.69\times10^{-4}\) instead of \(1.17\times10^{-5}\), yet the relative \(L^2\) error is only \(0.05\%\). In E13, the Klein--Gordon reaction coefficient is \(2.33\times10^{-4}\) instead of \(1.29\times10^{-3}\), with a relative \(L^2\) error of \(2.02\%\). Conversely, in E6 the transport and diffusion coefficients of the linear-flux conservation law are \(1.20\) and \(0.00306\) instead of \(1.14\) and \(0.00281\), and these small errors accumulate into a phase shift with relative \(L^2\) error \(24.7\%\). These cases suggest that errors in weak higher-order or reaction terms can leave the solution nearly unchanged over the observation window, whereas small errors in transport coefficients can dominate the reconstructed solution.

\subsection{Robustness to data noise}
\label{sec:noise_robustness}

We next evaluate robustness to Gaussian corruption of the augmented model input. Each input contains the solution field, first-order spatial-derivative, Fourier-real, and Fourier-imaginary channels. Corruption is applied after these channels have been constructed rather than by adding noise to a raw solution field and recomputing the derived features. Here, \(\eta\in\{0,0.02,0.05,0.1\}\) is the relative noise level: each entry \(z\) of the input is replaced by \(z+\eta|z|\xi\) with an independent \(\xi\sim\mathcal{N}(0,1)\), so the noise standard deviation equals \(100\eta\%\) of that entry's magnitude. We evaluate the Fourier-and-derivative model directly on these corrupted inputs.

\begin{figure}[htbp]
\centering
\includegraphics[width=\linewidth]{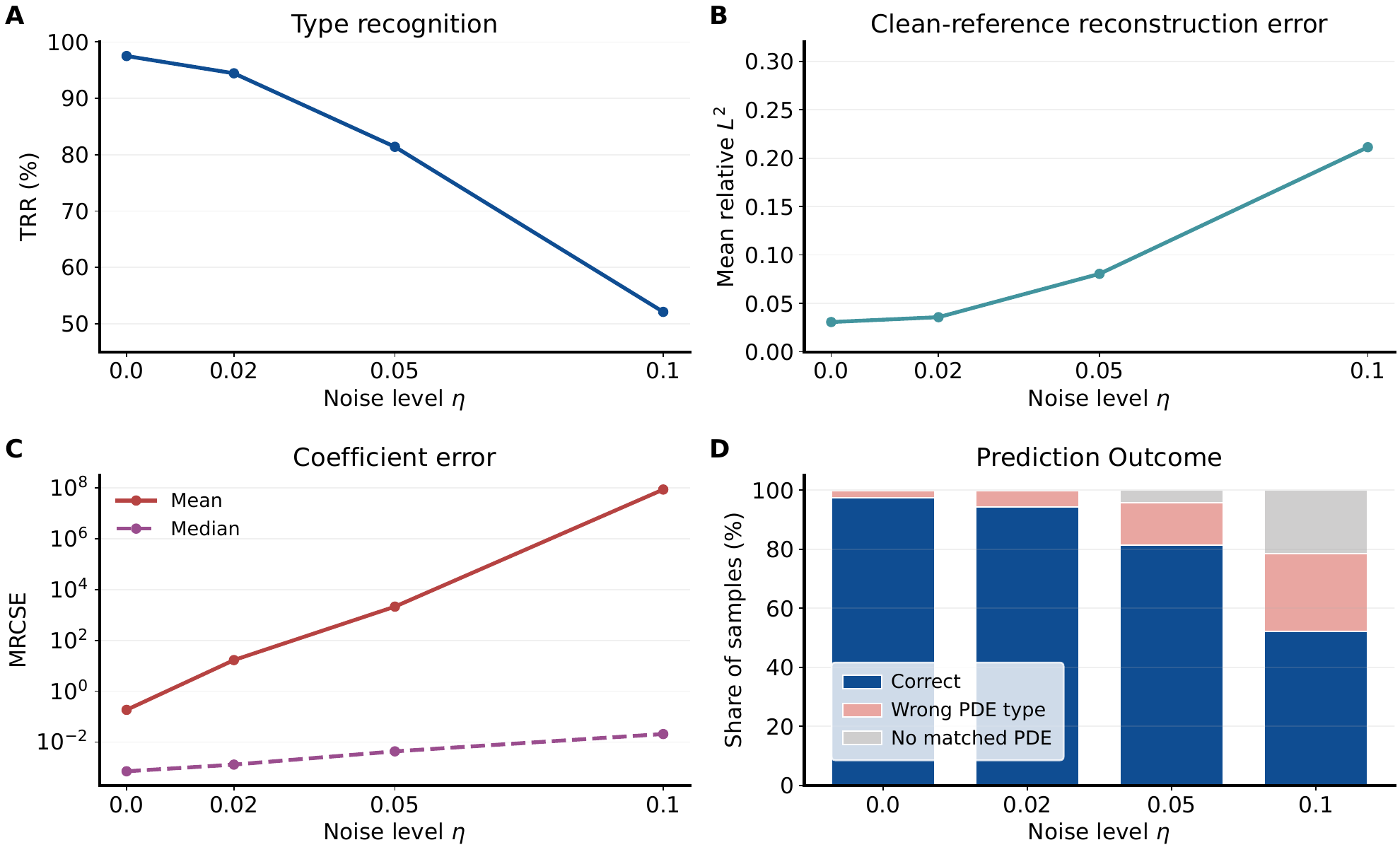}
\caption{\textbf{Corruption of the augmented input affects TRR, MRCSE, and relative \(L^2\) error at different rates.} The augmented tensor contains solution, first-order spatial-derivative, Fourier-real, and Fourier-imaginary channels, and \(\eta\) is the relative standard deviation of the multiplicative Gaussian noise applied to each entry. (\textbf{A}) TRR as \(\eta\) increases from clean input to \(0.1\). (\textbf{B}) Relative \(L^2\) error after reconstructing each decoded PDE from the clean initial condition and comparing it with the noise-free trajectory. (\textbf{C}) Mean and median MRCSE on a logarithmic scale; their growing separation indicates an increasingly heavy-tailed distribution. (\textbf{D}) Prediction categories over the fixed 5,120-example denominator, showing that wrong-PDE-type and no-matched-PDE outcomes become more frequent as corruption increases.}
\label{fig:noisy_summary}
\end{figure}

Figs.~\ref{fig:noisy_summary}A, B, and D report the aggregate effect of augmented-input corruption. The clean control has TRR \(97.50\%\), with a coefficient-setting-level 95\% bootstrap interval of \([96.54\%,98.30\%]\). TRR decreases to \(94.43\%\) \([92.64\%,96.02\%]\) at \(\eta=0.02\), \(81.39\%\) \([78.44\%,84.10\%]\) at \(\eta=0.05\), and \(52.09\%\) \([48.20\%,56.00\%]\) at \(\eta=0.1\). When decoded equations are reconstructed against the noise-free reference, the corresponding mean relative \(L^2\) errors are \(0.0308\), \(0.0358\), \(0.0805\), and \(0.211\), respectively. Their 95\% bootstrap intervals are \([0.0241,0.0384]\), \([0.0293,0.0435]\), \([0.0700,0.0924]\), and \([0.191,0.233]\). At \(\eta=0.02\), the TRR interval already lies below the clean-control interval, whereas the relative \(L^2\) intervals still overlap; both separate clearly from \(\eta=0.05\) onward. At \(\eta=0.1\), the prediction distribution also contains a substantially larger fraction of incorrect-type and no-matched-PDE outcomes.
Fig.~\ref{fig:noisy_summary}C provides a complementary view of the same experiment. Mean MRCSE grows faster than median MRCSE as \(\eta\) increases, indicating a heavy-tailed distribution. This distinction also appears in the bootstrap analysis: the median MRCSE interval at \(\eta=0.1\) is \([0.0130,0.0302]\), whereas the mean MRCSE interval is \([293,2.79\times10^{8}]\). Thus, when the symbolic type is still recovered, many predictions retain moderate MRCSE, but a small number of corrupted inputs dominate the mean. This again motivates reporting both robust and mean summaries rather than interpreting MRCSE through a single point estimate.

\begin{figure}[htbp]
\centering
\includegraphics[width=\linewidth]{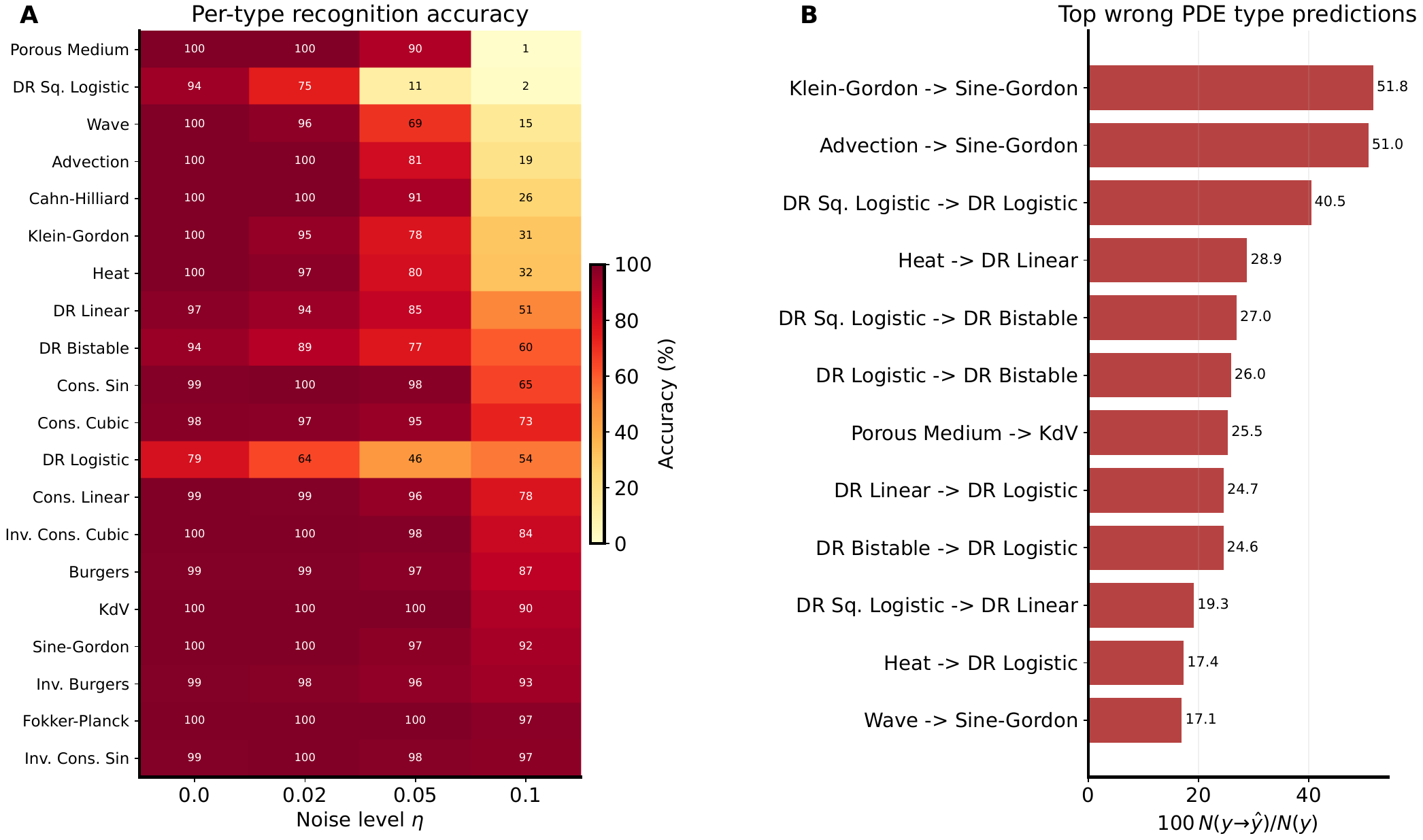}
\caption{\textbf{Sensitivity to augmented-input corruption depends on the PDE family.} (\textbf{A}) Per-type TRR as \(\eta\) increases. Fokker--Planck, Sine--Gordon, inviscid Burgers, and KdV retain high TRR at \(\eta=0.1\), whereas heat, wave, advection, Cahn--Hilliard, and several diffusion-reaction families show larger decreases. (\textbf{B}) Most frequent incorrect type transitions at \(\eta=0.1\). The $x$-axis reports \(100\,N(y\to\hat{y})/N(y)\), where \(N(y\to\hat{y})\) is the number of samples whose true PDE type \(y\) is predicted as \(\hat{y}\), and \(N(y)\) is the total number of samples with true type \(y\). The dominant transitions occur among related reaction-diffusion families and from heat and wave toward reaction or nonlinear-source families.}
\label{fig:noisy_type_breakdown}
\end{figure}

The per-type accuracy in Fig.~\ref{fig:noisy_type_breakdown}A shows that the aggregate trend is not representative of every PDE family. Sine--Gordon and Fokker--Planck maintain high TRR across the tested corruption levels, and inviscid Burgers changes little. In contrast, heat drops from complete recognition on clean input to \(31.9\%\) at \(\eta=0.1\), wave to \(15.2\%\), advection to \(19.0\%\), and Cahn--Hilliard to \(26.1\%\), and several diffusion-reaction families also show large decreases. Fig.~\ref{fig:noisy_type_breakdown}B shows that the most common transitions occur among related reaction-diffusion families and from heat and wave toward reaction or nonlinear-source families. These differences show that sensitivity to augmented-input corruption is strongly family dependent.

\begin{figure}[htbp]
\centering
\includegraphics[width=\linewidth]{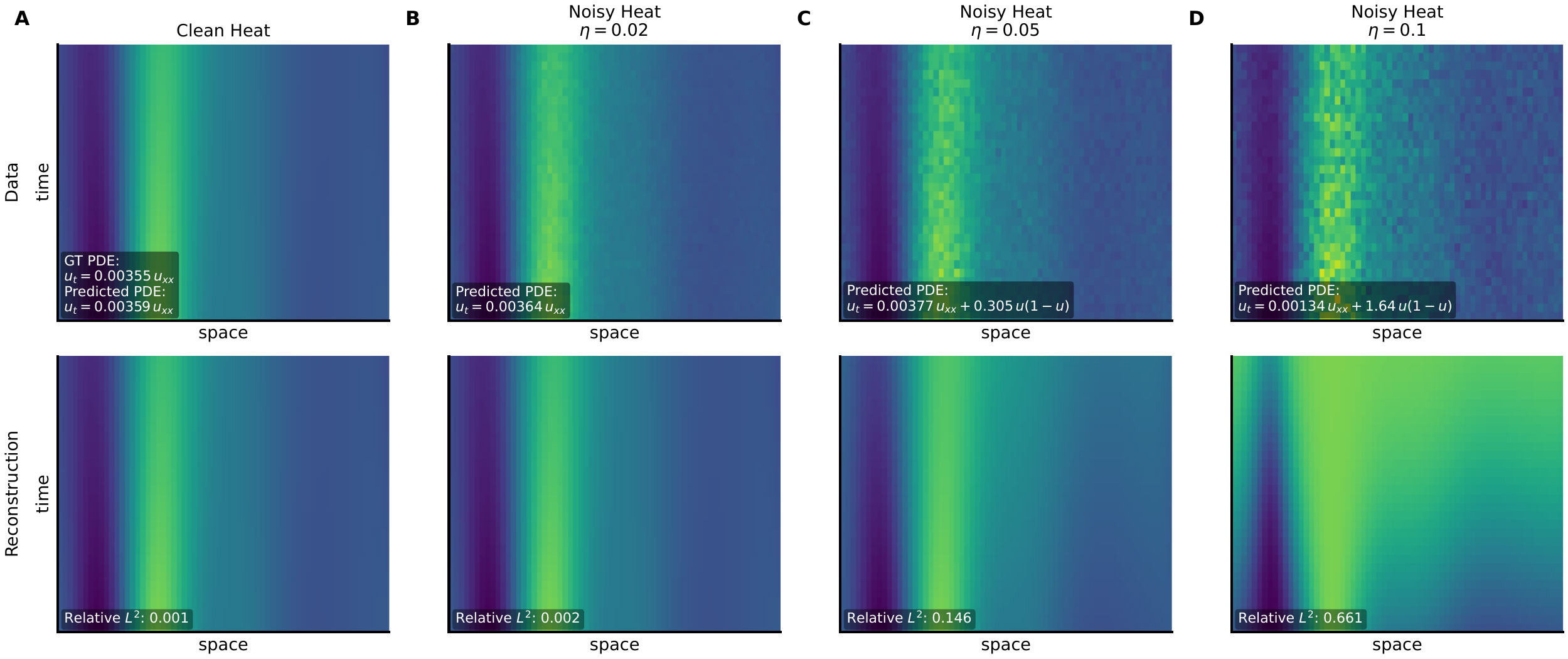}
\caption{\textbf{A corrupted heat example shows a spurious reaction term changing both the symbolic type and the reconstructed trajectory.} The top row visualizes the solution channel presented to the model at each corruption level. The bottom row reconstructs every decoded PDE from the same clean initial condition, and relative \(L^2\) is computed against the noise-free trajectory. (\textbf{A}) The clean input is recovered as a heat equation with relative \(L^2=0.001\). (\textbf{B}) At \(\eta=0.02\), the prediction remains a heat equation and gives relative \(L^2=0.002\). (\textbf{C}) At \(\eta=0.05\), the prediction adds a logistic reaction term, and the reconstruction has relative \(L^2=0.146\). (\textbf{D}) At \(\eta=0.1\), the logistic term grows and the relative \(L^2\) error reaches \(0.661\).}
\label{fig:noisy_heat_case_study}
\end{figure}

The heat-equation case study in Fig.~\ref{fig:noisy_heat_case_study} shows how corruption changes the decoded equation. The ground-truth clean PDE is \(u_t = 0.00355u_{xx}\), and the model predicts a close heat equation on the clean input and at \(\eta=0.02\), with relative \(L^2\) errors of \(0.001\) and \(0.002\). At \(\eta=0.05\) and \(\eta=0.1\), the prediction adds a logistic reaction term with coefficients \(0.305\) and \(1.64\), and reconstruction from the clean initial condition gives relative \(L^2\) errors of \(0.146\) and \(0.661\) against the noise-free trajectory. In this example, the spurious reaction term changes both the symbolic family and the reconstructed dynamics, so TRR and relative \(L^2\) error degrade together. The example is the lowest-index heat sample that is recognized correctly on clean input, misrecognized at \(\eta=0.05\), and decoded to a parseable PDE at every corruption level.

\subsection{Comparison with baselines}
\label{sec:baseline_comparison}

We next compare FoundPDE with representative symbolic discovery baselines, including SINDy~\cite{Brunton_2016} and its weak form~\cite{messenger2021weak}, PROSE-PDE~\cite{sun2025foundationmodelpartialdifferential}, and DISCOVER~\cite{PhysRevResearch.6.013182}. These experiments are diagnostic comparisons under the sparse-observation protocols used in this paper, not exhaustive benchmarks over all possible libraries, hyperparameters, resolutions, or per-equation tuning budgets. Sparse-regression and symbolic-search methods can be highly effective when the candidate library, derivative estimates, and optimization settings are well matched to the target equation. Our goal here is to evaluate how these baselines behave in the same low-resolution discovery setting where FoundPDE uses one pretrained model across many PDE families.

\paragraph{PROSE-PDE.}

We evaluate the pretrained PROSE-PDE model~\cite{sun2025foundationmodelpartialdifferential} on the same in-domain evaluation set, using only the solution data and removing the text input so that the comparison reflects data-only PDE identification. PROSE-PDE is a strong pretrained baseline: its mean relative \(L^2\) error is \(0.0345\) over 5,117 reconstructable predictions, close to the \(0.0308\) of FoundPDE over 5,115, showing that its predicted equations often reproduce the observed solutions well. However, the central objective of this paper is symbolic PDE recovery rather than solution fitting alone. Under that criterion, FoundPDE gives higher TRR, \(97.50\%\) compared with \(89.80\%\) for PROSE-PDE, and lower MRCSE, \(0.185\) over 4,992 evaluable predictions compared with \(0.905\) over 4,598. The advantage is especially visible in families where the symbolic form is easy to confuse despite similar solution behavior: FoundPDE improves over PROSE-PDE on Klein--Gordon (\(99.5\%\) versus \(63.5\%\)), wave (\(100\%\) versus \(85.0\%\)), square-logistic reaction-diffusion (\(93.7\%\) versus \(67.7\%\)), and logistic reaction-diffusion (\(79.0\%\) versus \(57.7\%\)). These results suggest that FoundPDE's training objective and output representation provide a stronger prior for symbolic family recognition and numerical calibration, while PROSE-PDE remains close in relative \(L^2\) error.

\paragraph{SINDy.}

We also compare against SINDy~\cite{Brunton_2016,doi:10.1126/sciadv.1602614} and its weak form~\cite{messenger2021weak} on four diffusion-reaction equations with increasingly nonlinear reaction terms. This benchmark is narrower than the DISCOVER comparison. Every method receives the same \(32\times64\) trajectories: the first initial condition of each equation, which is also the instance used for DISCOVER, and, for FoundPDE and both SINDy variants, the 19 remaining initial conditions of the same equation. Both SINDy variants use the candidate library \(\{1,u,u^2,u^3,u^4,u_x,u_{xx}\}\), which contains every true term, so the comparison isolates term selection and coefficient estimation from library design. Sparsity is selected by the Akaike information criterion on held-out time steps~\cite{mangan2017model} for SINDy and by the weak-SINDy model-selection loss~\cite{messenger2021weak} for weak SINDy; Appendix~\ref{app:sindy_protocol} gives the full protocol.

Table~\ref{tab:reaction_results} reports the recovered equations for the first initial condition together with the number of correctly identified initial conditions. On clean data, both SINDy variants are competitive: with the true terms available, they recover three of the four displayed instances, and their coefficients are closer to the ground truth than those of FoundPDE whenever the structure is correct. Over all 80 trajectories, FoundPDE identifies 68 correctly, compared with 50 for SINDy and 53 for weak SINDy. The methods fail on different equations: both SINDy variants most often miss the logistic reaction, either adding higher-order terms or omitting the small diffusion term, whereas the errors of FoundPDE concentrate on the bistable and square-logistic reactions, which it most often predicts as linear reaction. The difference becomes large under noise (Appendix Table~\ref{tab:sindy_noise}). When multiplicative Gaussian noise of relative level \(\eta\) is added to the observed solution and the FoundPDE input features are recomputed from the noisy field, FoundPDE identifies 54 of the 80 trajectories at \(\eta=0.01\) and 45 at \(\eta=0.02\), whereas SINDy recovers at most 9 and weak SINDy at most 17 across the derivative and test-function settings we tested. On this grid, derivative estimates from noisy data are dominated by noise, and SINDy typically drops the diffusion term in favor of spurious polynomial terms. This comparison covers four equations and is not an exhaustive benchmark of sparse regression; it indicates that the pretrained data-to-symbol prior is most useful when derivatives cannot be estimated reliably.

\begin{table}[htbp]
\caption{\textbf{Sparse regression is competitive on clean diffusion-reaction data when its library contains the true terms.} All test PDEs have the form \(u_t = \nu u_{xx} + \rho R(u)\) and are observed on the \(32\times64\) grid. Each entry is the equation recovered from the first initial condition, the instance also used for DISCOVER; parentheses give the number of the 20 initial conditions of the same equation for which the recovered terms match the ground truth. FoundPDE entries are raw decoded equations before BFGS refinement. SINDy and weak SINDy use the library \(\{1,u,u^2,u^3,u^4,u_x,u_{xx}\}\), and their polynomial reaction terms are shown in expanded form. \textcolor{green!55!black}{Green} entries match the ground-truth structure, while \textcolor{red}{red} entries indicate a structural mismatch. DISCOVER was run on the first initial condition only. Appendix Table~\ref{tab:sindy_noise} reports the same comparison under noise.}
\label{tab:reaction_results}
\centering
\renewcommand{\arraystretch}{1.25}
\setlength{\tabcolsep}{6pt}
\resizebox{\textwidth}{!}{%
\begin{tabular}{@{}lllll@{}}
\toprule
Ground truth & FoundPDE & SINDy & Weak SINDy & DISCOVER \\
\midrule
\(u_t = 0.000802u_{xx} + 0.0472u\)
& \(u_t = 0.000722u_{xx} + \textcolor{green!55!black}{0.0492u}\) (19/20)
& \(u_t = 0.000799u_{xx} + \textcolor{green!55!black}{0.0470u}\) (16/20)
& \(u_t = 0.000792u_{xx} + \textcolor{green!55!black}{0.0472u}\) (20/20)
& \(u_t = 0.0008u_{xx} + \textcolor{green!55!black}{0.0486u}\) \\
\addlinespace
\(u_t = 0.00228u_{xx} + 0.794u(1-u)\)
& \(u_t = 0.00231u_{xx} + \textcolor{green!55!black}{0.813u(1-u)}\) (20/20)
& \splitcell{\(u_t = 0.00228u_{xx} + \textcolor{red}{0.778u - 0.729u^2}\)\\\(\quad\textcolor{red}{{}-0.084u^3 + 0.034u^4}\) (6/20)}
& \(u_t = \textcolor{red}{0.962u - 1.024u^2}\) (3/20)
& \(u_t = 0.0034u_{xx} + \textcolor{green!55!black}{0.0309u(1-u)}\) \\
\addlinespace
\(u_t = 0.00496u_{xx} + 0.239u^2(1-u)\)
& \(u_t = 0.00447u_{xx} + \textcolor{red}{0.529u^2(1-u)^2}\) (13/20)
& \(u_t = 0.00492u_{xx} + \textcolor{green!55!black}{0.248u^2 - 0.253u^3}\) (16/20)
& \(u_t = 0.00489u_{xx} + \textcolor{green!55!black}{0.241u^2 - 0.242u^3}\) (20/20)
& \(u_t = 0.001u_{xx} + \textcolor{red}{0.0428u + 0.005}\) \\
\addlinespace
\(u_t = 0.00344u_{xx} + 0.125u^2(1-u)^2\)
& \(u_t = 0.00325u_{xx} + \textcolor{green!55!black}{0.171u^2(1-u)^2}\) (16/20)
& \splitcell{\(u_t = 0.00345u_{xx} + \textcolor{green!55!black}{0.125u^2}\)\\\(\quad\textcolor{green!55!black}{{}-0.249u^3 + 0.125u^4}\) (12/20)}
& \splitcell{\(u_t = 0.00343u_{xx} + \textcolor{green!55!black}{0.126u^2}\)\\\(\quad\textcolor{green!55!black}{{}-0.252u^3 + 0.126u^4}\) (10/20)}
& \(u_t = 0.0034u_{xx} + \textcolor{red}{0.0333u(1-u)}\) \\
\midrule
Correct over 80 trajectories & 68 & 50 & 53 & -- \\
\bottomrule
\end{tabular}%
}
\end{table}

\paragraph{DISCOVER.}

Like SINDy, DISCOVER~\cite{PhysRevResearch.6.013182} is a per-instance method, but it performs symbolic search guided by reinforcement learning rather than sparse regression: for each target trajectory, the recovery program is run independently rather than reusing a single model across PDE families. For the sparse-resolution evaluation, trajectories are taken from the existing FoundPDE evaluation corpus, which stores 20 initial-condition realizations for each completed PDE. We select the first trajectory from each reported PDE block and run DISCOVER separately on that instance. Each run is trained for 100 epochs with 50000 samples, batch size 500, and \(\epsilon=0.2\). The candidate vocabulary is configured for each target equation to include the symbolic primitives appearing in its analytical form; \texttt{sin} and \texttt{cos} are added for equations with trigonometric terms. This target-informed construction isolates expression search from operator-vocabulary selection. Because DISCOVER can return expressions whose displayed forms differ despite algebraic equivalence, we manually verify the displayed predictions after simplification before assigning their structural categories. Table~\ref{tab:discover} reports these DISCOVER results together with dense-resolution DISCOVER runs and representative FoundPDE predictions from the corresponding PDE families. Because the dense-resolution and FoundPDE entries use separately selected coefficient instances, the comparison is made at the family level rather than as a sample-matched coefficient benchmark.

\begin{table}[htbp]
\caption{\textbf{A family-level representative comparison shows more consistent symbolic-family recovery by FoundPDE than by per-instance DISCOVER.} The ground-truth column gives the target of the sparse \(32\times64\) DISCOVER run. The \(128\times128\) DISCOVER entry is a separately instantiated representative of the same PDE family, and the FoundPDE entry is selected from the no-noise \(32\times64\) evaluation set using the same family and the nearest available coefficient setting. The columns therefore compare recovery behavior across shared PDE families, not predictions on identical coefficients and initial conditions, and coefficient values should not be compared directly across columns. DISCOVER is optimized independently for every displayed instance, whereas all FoundPDE entries are produced by the same pretrained model without per-instance optimization. \textcolor{green!55!black}{Green} indicates recovery of the target symbolic family for the corresponding instance, \textcolor{relatedorange}{orange} indicates an algebraically related or simplifiable expression with a non-canonical or incomplete form, and \textcolor{red}{red} indicates a structural mismatch.}
\label{tab:discover}
\centering
\renewcommand{\arraystretch}{1.12}
\setlength{\tabcolsep}{4pt}
\resizebox{\textwidth}{!}{%
\begin{tabular}{@{}llll@{}}
\toprule
Ground truth & DISCOVER \(32\times64\) & DISCOVER \(128\times128\) & FoundPDE \(32\times64\) \\ \midrule
\splitcell{\textbf{Burgers}\\$u_t = 0.00458u_{xx}-0.697uu_x$}
& \relatedcell{$u_t = 0.0023u_{xx}-0.3596(u^2)_x$}
& \relatedcell{$u_t = 0.0056u_{xx}-0.8332(u^2)_x$}
& \goodcell{$u_t = 0.00357u_{xx}-0.533uu_x$} \\
\splitcell{\textbf{Heat}\\$u_t = 0.00409u_{xx}$}
& \relatedcell{\splitcell{$u_t = 0.0164u_{xx}$\\$/((x^2)_{xx})^3$}}
& \relatedcell{$u_t = 0.0053u_{xx}/(x^2)_{xx}$}
& \goodcell{$u_t = 0.0041u_{xx}$} \\
\splitcell{\textbf{Advection}\\$u_t = -0.562u_x$}
& \relatedcell{$u_t = -0.092(x^3)_{xxx}u_x$}
& \relatedcell{$u_t = -0.0594(x^3)_{xxx}u_x$}
& \goodcell{$u_t = -0.775u_x$} \\
\splitcell{\textbf{Cahn--Hilliard}\\$u_t = -2.3875{\times}10^{-6}u_{xxxx}$\\$-1.5u_xu_x-1.5uu_{xx}$}
& \badcell{\splitcell{$u_t = 0.0089u_{xx}-0.0077u(u_{xx}+u_{xx})$\\$-2.9985u^2+1.7466u$}}
& \badcell{\splitcell{$u_t = -0.0056uu_{xx}+0.0028u_{xx}+0.9546u$\\$-0.8303u(u+u)$}}
& \goodcell{\splitcell{$u_t = -2.39{\times}10^{-6}u_{xxxx}$\\$-1.5u_xu_x-1.5uu_{xx}$}} \\
\splitcell{\textbf{Conservation (cubic flux)}\\$u_t = -0.341u^2u_x+0.00057u_{xx}$}
& \relatedcell{\splitcell{$u_t = 0.0009(u^2)_{xx}$\\$-0.1176(uuu)_x$}}
& \relatedcell{\splitcell{$u_t = -0.454(u^3)_x$\\$+0.003(u^3)_{xx}$}}
& \goodcell{$u_t = -1.08u^2u_x+0.000544u_{xx}$} \\
\splitcell{\textbf{Conservation (linear flux)}\\$u_t = -0.98u_x+0.00267u_{xx}$}
& \relatedcell{$u_t = -0.165u_x(x^3)_{xxx}$}
& \relatedcell{$u_t = -0.844u_x$}
& \goodcell{$u_t = -1.12u_x+0.00257u_{xx}$} \\
\splitcell{\textbf{Conservation (sine flux)}\\$u_t = -0.388\cos u\,u_x+0.00205u_{xx}$}
& \relatedcell{$u_t = -0.4032\cos u\,u_x$}
& \relatedcell{$u_t = -1.1314\cos u\,u_x$}
& \goodcell{\splitcell{$u_t = -0.289\cos u\,u_x$\\$+0.00279u_{xx}$}} \\
\splitcell{\textbf{Diffusion reaction (bistable)}\\$u_t = 0.00496u_{xx}+0.239u^2(1-u)$}
& \badcell{\splitcell{$u_t = 0.0428u$\\$+0.005\frac{(u_{xx})^2+u_{xx}}{u_{xx}}$}}
& \relatedcell{\splitcell{$u_t = 0.7769(u+u)^2$\\$-1.037u(u^2+2u)$}}
& \goodcell{$u_t = 0.00423u_{xx}+0.197u(1-u)(u-0)$} \\
\splitcell{\textbf{Diffusion reaction (linear)}\\$u_t = 0.000802u_{xx}+0.0472u$}
& \relatedcell{\splitcell{$u_t = 0.0486u$\\$+0.0008\frac{u_{xx}^2}{u_{xx}}$}}
& \goodcell{$u_t = 0.167u+0.0013u_{xx}$}
& \goodcell{$u_t = 0.000303u_{xx}+0.0357u$} \\
\splitcell{\textbf{Diffusion reaction (logistic)}\\$u_t = 0.00228u_{xx}+0.794u(1-u)$}
& \badcell{\splitcell{$u_t = -0.1511(u^2+u)(u+u)$\\$+0.0029u_{xx}+0.6085u$}}
& \goodcell{$u_t = 1.1724u-1.1692u^2+0.0049u_{xx}$}
& \goodcell{$u_t = 0.00228u_{xx}+0.827u(1-u)$} \\
\splitcell{\textbf{Diffusion reaction (square logistic)}\\$u_t = 0.00344u_{xx}+0.127u^2(1-u)^2$}
& \badcell{\splitcell{$u_t = 0.0309u-0.0111u(u+u+u)$\\$+0.0034u_{xx}$}}
& \relatedcell{\splitcell{$u_t = 1.5417(u+u^2)^2$\\$-6.1677u^3+0.0048u_{xx}$}}
& \goodcell{$u_t = 0.00368u_{xx}+0.103u^2(1-u)^2$} \\
\splitcell{\textbf{Fokker--Planck}\\$u_t = (4.54\sin(10x)u)_x$\\$+0.3765u_{xx}$}
& \badcell{\splitcell{$u_t = -966.9506$\\$(\sin^3(x(x+u))\sin(xu))^3$}}
& \badcell{$u_t = -9847.0103(u(x^2)_{xx})^4$}
& \goodcell{\splitcell{$u_t = (2.67\sin(10x)u)_x$\\$+0.2215u_{xx}$}} \\
\splitcell{\textbf{Inviscid Burgers}\\$u_t = -1.4uu_x$}
& \relatedcell{\splitcell{$u_t = -0.0002(u^4x)_{xxx}$\\$-0.661(u^2)_x$}}
& \relatedcell{$u_t = 0.0022(u^4)_{xx}-0.8151(u^2)_x$}
& \goodcell{$u_t = -1.42uu_x$} \\
\splitcell{\textbf{Inviscid conservation (cubic flux)}\\$u_t = -0.483u^2u_x$}
& \relatedcell{\splitcell{$u_t = -0.1682(u^3)_x$\\$+0.0021(u^2)_{xx}$}}
& \relatedcell{\splitcell{$u_t = 0.0009(u^2)_{xx}$\\$-0.1288(u^3)_x$}}
& \goodcell{$u_t = -0.374u^2u_x$} \\
\splitcell{\textbf{Inviscid conservation (sine flux)}\\$u_t = -1.15\cos u\,u_x$}
& \badcell{\splitcell{$u_t = -0.0004\cos(u^3)(u_{xxx}+u_{xx})$\\$-0.8614u_x$}}
& \badcell{$u_t = -1.0036\cos(u^4)u_x$}
& \goodcell{$u_t = -1.15\cos u\,u_x$} \\
\splitcell{\textbf{KdV}\\$u_t = -0.00084u_{xxx}-0.5uu_x$}
& \relatedcell{\splitcell{$u_t = -0.1861(u^2)_x$\\$-0.0003(u^2)_{xxx}$}}
& \relatedcell{$u_t = -0.0001u_{xxx}-0.2339(u^2)_x$}
& \goodcell{$u_t = -0.00077u_{xxx}-0.5uu_x$} \\
\splitcell{\textbf{Porous medium}\\$u_t = 0.05(u^2)_{xx}$}
& \relatedcell{$u_t = 0.0997(u^2)_{xx}/(x^2)_{xx}$}
& \goodcell{$u_t = 0.0504(u^3)_{xx}$}
& \goodcell{$u_t = 0.05(u^2)_{xx}$} \\
\bottomrule
\end{tabular}%
}
\end{table}

Table~\ref{tab:discover} highlights three practical differences between per-instance symbolic search and amortized data-to-symbol prediction. First, DISCOVER performs inference one instance at a time: every entry requires a separate optimization run, whereas FoundPDE applies one pretrained model across the listed families. Second, the family-level results indicate that increased resolution alone does not produce uniformly reliable DISCOVER recovery. Several dense-resolution representatives are close to the intended family, especially for linear and logistic diffusion--reaction equations and porous-medium diffusion, while Cahn--Hilliard, Fokker--Planck, and sine-flux conservation remain non-canonical or structurally incorrect. Third, DISCOVER can produce complicated expressions whose simplified form contains a useful operator. For example, \((u^2)_x\) exposes Burgers-type nonlinear transport, \((x^3)_{xxx}u_x\) reduces to an advection term, and the porous-medium expression normalizes through \((x^2)_{xx}\). These orange entries are therefore not complete failures, but they require algebraic simplification and can still omit terms needed for the target family. The representative FoundPDE entries recover the corresponding symbolic family in every displayed row.

\subsection{Parameter-OOD generalization}
\label{sec:ood_parameter}

In-domain evaluation measures generalization across independently sampled coefficients and initial conditions within the support used during pretraining. It does not establish whether the model can extrapolate beyond that coefficient support. We therefore construct a strict parameter-OOD benchmark in which the symbolic PDE families remain unchanged, while at least one physical coefficient lies outside the complete in-domain range. This setting isolates coefficient extrapolation from structural generalization: the model has encountered each operator family during pretraining, but not the tested parameter values.

The OOD dataset contains 5,120 trajectories from 19 PDE families, comprising 640 coefficient-instantiated equations with eight independently sampled initial conditions per equation. The initial conditions are drawn from the same family-specific distributions used for pretraining so that the distribution shift lies in the PDE coefficients. For equations with multiple coefficients, one coefficient is moved outside the in-domain support while the remaining coefficients are held at their reference values. The exceptions are Burgers and the viscous cubic-flux conservation law, in which the viscosity \(\epsilon\) is increased with the flux coefficient \(k\) to keep \(k\le180\epsilon\) for numerical stability. The pretrained Fourier-and-derivative model is evaluated without fine-tuning on \(32\times64\) temporal--spatial grids, using the family-specific domains and final times of the in-domain corpus. No additional evaluation-time noise is injected. Sixty-two samples produce empty decoder outputs and 30 produce expressions that cannot be assigned a PDE type; all are conservatively counted as type-recognition failures. Porous-medium equations are excluded because changing the discrete exponent \(m\) alters the symbolic structure of \(u_t=(u^m)_{xx}\) rather than only a continuous parameter. Appendix Table~\ref{tab:ood_parameter_ranges} reports the strictly disjoint coefficient supports and their stability restrictions.

\begin{figure}[htbp]
\centering
\includegraphics[width=\linewidth]{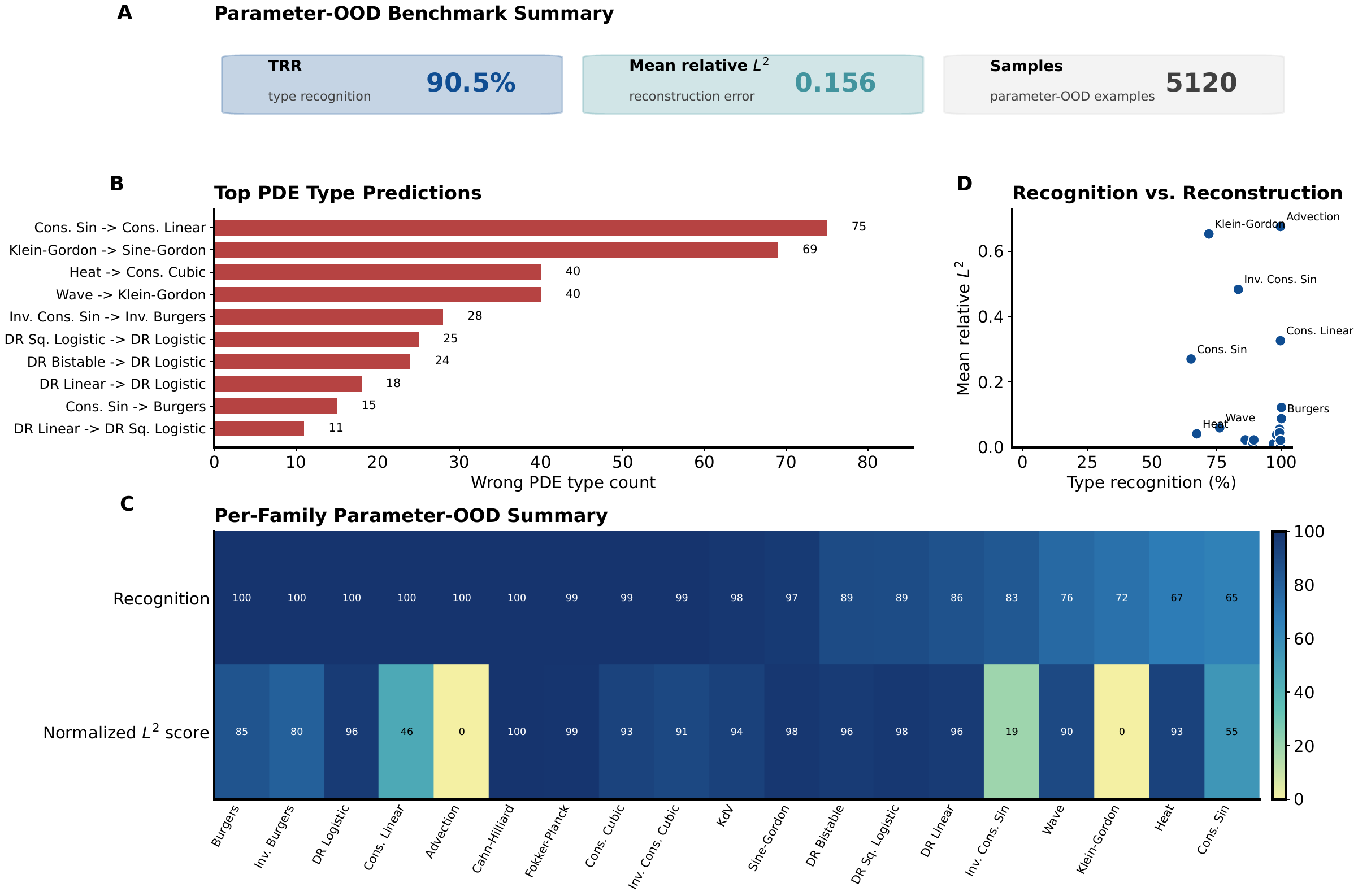}
\caption{\textbf{Strict parameter-OOD evaluation reveals a moderate aggregate decrease and concentrated family-level failures.} (\textbf{A}) Aggregate TRR and mean relative \(L^2\) error over 5,120 samples whose coefficient supports are disjoint from the in-domain ranges. (\textbf{B}) The ten most frequent incorrect type transitions. (\textbf{C}) Per-family TRR and normalized relative \(L^2\) score, computed as \(100\,\mathrm{clip}(1-\mathrm{Rel}L^2/0.60,0,1)\) for visualization. (\textbf{D}) Per-family TRR versus mean relative \(L^2\) error. Advection and viscous linear conservation illustrate that high structural recognition does not necessarily imply low relative \(L^2\) error.}
\label{fig:ood_parameter_summary}
\end{figure}

FoundPDE correctly identifies 4,635 of the 5,120 equations, giving a strict parameter-OOD TRR of \(90.53\%\) (Fig.~\ref{fig:ood_parameter_summary}A), with a coefficient-setting-level 95\% bootstrap interval of \([89.22\%,91.80\%]\). The comparable 19-family in-domain TRR is \(97.34\%\), so moving every evaluated coefficient vector outside the in-domain support reduces TRR by 6.81 percentage points. Among the 5,011 reconstructable predictions, the raw decoded equations give a mean relative \(L^2\) error of \(0.156\), with a 95\% interval of \([0.139,0.174]\), a median of \(0.0347\), and a 95th percentile of \(0.777\). MRCSE is evaluable for 4,630 type-matched predictions. Its mean is \(0.228\), with a 95\% interval of \([0.140,0.331]\), whereas its median and 95th percentile are \(0.00458\) and \(0.275\), respectively; the median MRCSE interval is \([0.00395,0.00544]\). The largest single error comes from a heat-equation prediction whose decoded diffusion coefficient is \(0.063\) instead of \(0.00708\), producing an MRCSE of \(62.4\). We report the mean together with robust distributional statistics rather than using the mean alone to characterize numerical calibration.

The family-level values in Table~\ref{tab:strict_ood_results} show that the aggregate decrease is concentrated rather than uniform. Burgers and inviscid Burgers reach \(100\%\) TRR, while advection, logistic diffusion--reaction, viscous linear conservation, Cahn--Hilliard, Fokker--Planck, viscous and inviscid cubic conservation, KdV, and Sine--Gordon all exceed \(96\%\). The cubic-flux results are particularly useful because they test nonlinear conservation operators under strict coefficient shifts: the viscous and inviscid variants reach \(99.26\%\) and \(99.24\%\) TRR, respectively.

\begin{table}[htbp]
\caption{\textbf{Strict parameter-OOD performance is strong for most pretrained operator families but degrades for sine-flux conservation and for large coefficient shifts in heat, Klein--Gordon, and wave.} TRR includes empty decoder outputs as failures. Mean relative \(L^2\) error is computed on 5,011 reconstructable predictions. Median MRCSE is computed on 4,630 type-matched predictions with extractable packed coefficients. Both numerical metrics use the raw decoded equations before BFGS refinement. The final column reports the most frequent parsed incorrect type; ``none observed'' indicates that the only failures, if any, are empty decoder outputs.}
\label{tab:strict_ood_results}
\centering
\scriptsize
\setlength{\tabcolsep}{3pt}
\begin{tabular}{@{}>{\raggedright\arraybackslash}p{0.27\linewidth}cccc>{\raggedright\arraybackslash}p{0.23\linewidth}@{}}
\toprule
Family & Samples & TRR (\%) $\uparrow$ & Mean Rel. \(L^2\) $\downarrow$ & Median MRCSE $\downarrow$ & Most frequent incorrect type \\
\midrule
Burgers & 272 & 100.00 & 0.088 & 0.008 & None observed \\
Inviscid Burgers & 272 & 100.00 & 0.122 & 0.005 & None observed \\
Advection & 272 & 99.63 & 0.676 & 2.568 & None observed \\
Diffusion--reaction, logistic & 272 & 99.63 & 0.021 & 0.007 & None observed \\
Viscous conservation, linear flux & 272 & 99.63 & 0.326 & 0.011 & Sine-flux conservation \\
Cahn--Hilliard & 264 & 99.62 & 0.000 & 0.003 & None observed \\
Fokker--Planck & 272 & 99.26 & 0.006 & 0.001 & None observed \\
Viscous conservation, cubic flux & 272 & 99.26 & 0.044 & 0.002 & Burgers \\
Inviscid conservation, cubic flux & 264 & 99.24 & 0.055 & 0.002 & Inviscid Burgers \\
KdV & 272 & 98.16 & 0.038 & 0.008 & Heat \\
Sine--Gordon & 264 & 96.97 & 0.011 & 0.001 & None observed \\
Diffusion--reaction, bistable & 272 & 89.34 & 0.022 & 0.004 & Logistic reaction \\
Diffusion--reaction, square-logistic & 272 & 88.97 & 0.013 & 0.004 & Logistic reaction \\
Diffusion--reaction, linear & 272 & 86.03 & 0.022 & 0.005 & Logistic reaction \\
Inviscid conservation, sine flux & 264 & 83.33 & 0.483 & 0.040 & Inviscid Burgers \\
Wave & 264 & 76.14 & 0.059 & 0.001 & Klein--Gordon \\
Klein--Gordon & 264 & 71.97 & 0.653 & 0.008 & Sine--Gordon \\
Heat & 272 & 67.28 & 0.041 & 0.013 & Cubic-flux conservation \\
Viscous conservation, sine flux & 272 & 65.07 & 0.270 & 0.015 & Linear conservation \\
\bottomrule
\end{tabular}
\end{table}

The principal structural errors occur among closely related operator families and for families whose coefficients are moved far beyond the training range (Fig.~\ref{fig:ood_parameter_summary}B). Viscous sine-flux conservation reaches \(65.07\%\) TRR, with 75 predictions mapped to linear-flux conservation. Heat reaches \(67.28\%\). Every heat test equation has a diffusion coefficient above the in-domain support: at least seven of the eight initial conditions are recognized for every equation with a diffusion coefficient below \(0.0065\), but recognition falls to as few as two of eight between \(0.0068\) and \(0.0075\), and 40 predictions are mapped to cubic-flux conservation. Klein--Gordon (\(71.97\%\)) and wave (\(76.14\%\)) degrade in the same way at their largest coefficient shifts, with 69 Klein--Gordon samples predicted as Sine--Gordon and 40 wave samples predicted as Klein--Gordon. Appendix~\ref{app:convergence} shows that the same model trained with a smaller budget extrapolates more reliably for these three families but less reliably for the reaction families, so the extrapolation profile depends on the training budget. The diffusion--reaction families reach \(89.34\%\) (bistable), \(88.97\%\) (square-logistic), and \(86.03\%\) (linear), with logistic reaction the most frequent incorrect type in each case. The sensitivity also depends on which coefficient is shifted. Bistable recognition is \(86.03\%\) for the reaction-coefficient tail and \(92.65\%\) for the diffusion-coefficient tail. Square-logistic shows a similar contrast of \(83.82\%\) and \(94.12\%\), whereas logistic reaction remains above \(99\%\) for both shifts.

These reaction-family errors often retain small relative \(L^2\) errors. The family means are \(0.022\), \(0.022\), and \(0.013\) for bistable, linear, and square-logistic reaction, respectively, despite their lower TRR. For the 24 bistable samples predicted as logistic reaction, the median relative \(L^2\) error is \(0.033\) over the observed time window. This observation does not establish mathematical indistinguishability between the operators. It shows that low relative \(L^2\) error over a finite observation window is insufficient evidence of correct symbolic recovery for related reaction terms.

Advection provides the clearest separation between structural and numerical generalization (Figs.~\ref{fig:ood_parameter_summary}C and~D). Its TRR is \(99.63\%\), but every parsed prediction from the negative-velocity tail has the wrong coefficient sign before refinement. All positive-tail predictions have the correct sign, with a median MRCSE of \(0.0136\), whereas the negative-tail median is \(3.115\). Consequently, advection has a mean relative \(L^2\) error of \(0.676\) despite nearly perfect type recognition. For \(u_t+\beta u_x=0\), an error in \(\beta\) produces a spatial phase displacement that accumulates with time, while a sign error reverses the propagation direction. The result therefore identifies a directional coefficient-extrapolation failure that TRR alone cannot reveal.

Viscous linear conservation exhibits a related separation. Its TRR is \(99.63\%\), but its mean relative \(L^2\) error is \(0.326\), increasing to approximately \(0.52\) when the transport coefficient is shifted. Conversely, some strict TRR failures are numerically minor. One wave prediction adds a Klein--Gordon reaction term with coefficient \(0.0031\) and obtains a relative \(L^2\) error of \(0.0028\). Together, these cases show why TRR, MRCSE, and relative \(L^2\) error must be interpreted separately. The strict benchmark supports coefficient extrapolation across most pretrained PDE families, while identifying signed transport, sine-flux conservation, and large coefficient shifts in heat, Klein--Gordon, and wave as the principal limitations under the tested ranges.

\subsection{Few-shot fine-tuning}
\label{sec:few_shot_results}

We also evaluate whether FoundPDE pretraining can support adaptation to new PDE families using limited additional data. All four fine-tuning trajectories begin from the separately pretrained 100-epoch, 512,000-sample checkpoint defined in Sec.~\ref{sec:few_shot_finetuning}. Fine-tuning epoch 0 is the state before any target-family update. This study considers the four held-out composite families defined in Sec.~\ref{sec:few_shot_finetuning} (advection--Burgers, advection--cubic-Burgers, advection--sine, and Burgers--cubic-Burgers) and compares pretrained adaptation with training from scratch. For the advection--Burgers fine-tuning run, we additionally measure retention on the original pretraining families.

\begin{figure}[htbp]
\centering
\includegraphics[width=\linewidth]{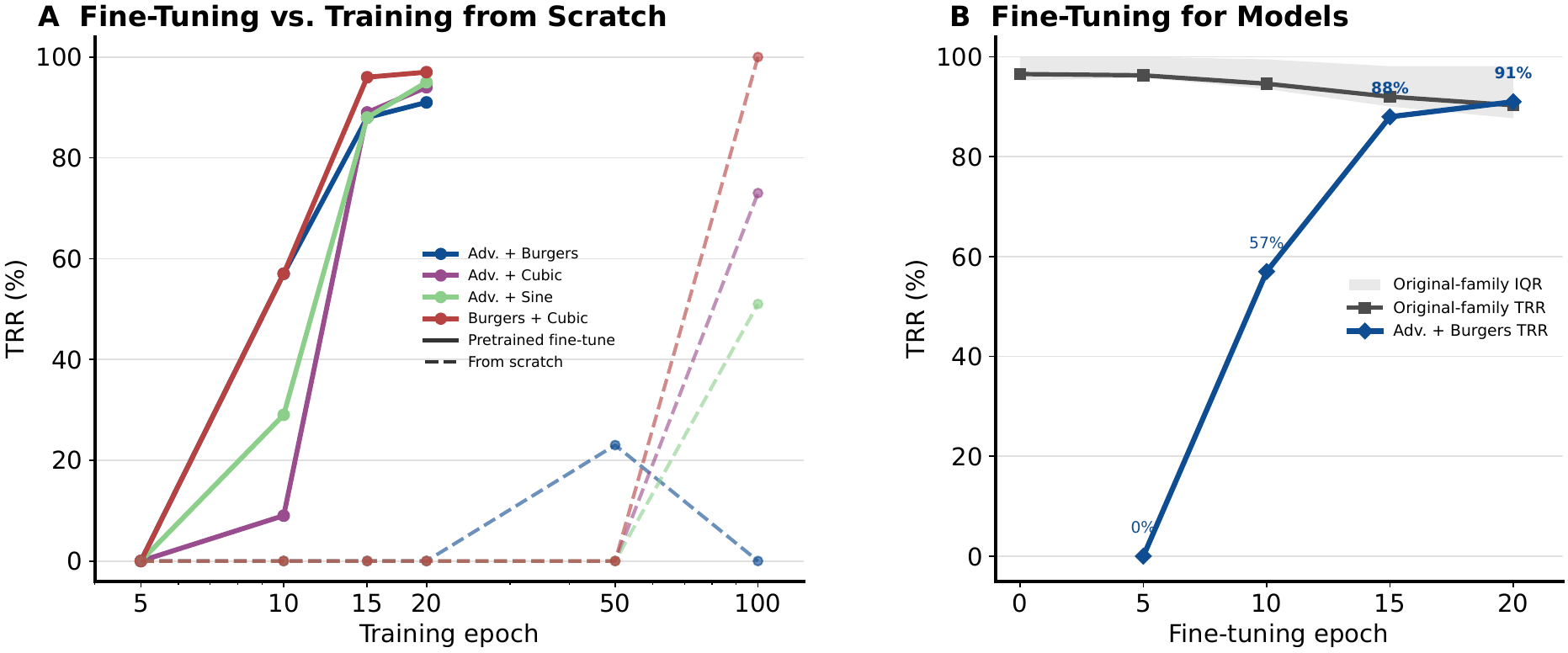}
\caption{\textbf{Single-family checkpoint adaptation reaches high composite recovery within 20 epochs, while joint four-family scratch training requires a longer horizon.} (\textbf{A}) Separate pretrained fine-tuning runs (solid curves) versus one jointly trained scratch model (dashed curves). Both protocols use 10 optimizer updates per epoch, but their batch composition differs. By epoch 20, fine-tuning reaches \(91.0\%\) exact recovery on advection--Burgers, \(94.0\%\) on advection--cubic-Burgers, \(95.0\%\) on advection--sine, and \(97.0\%\) on Burgers--cubic-Burgers. At epoch 100, scratch training remains below these values on three families but reaches \(100\%\) on Burgers--cubic-Burgers. The curves are single-run trajectories, and epochs 20 and 100 are fixed comparison horizons rather than test-selected checkpoints. (\textbf{B}) Retention along the advection--Burgers fine-tuning run: target-family TRR increases rapidly, while the fixed-denominator macro-average TRR across the 20 original families decreases from \(96.52\%\) to \(88.50\%\). Missing outputs are counted as failures.}
\label{fig:fewshot_transfer_summary}
\end{figure}

Fig.~\ref{fig:fewshot_transfer_summary}A reports the adaptation trajectories for the four composite targets. At epoch 5, none of the fine-tuned models achieves exact structural recovery, but several already recover a substantial fraction of the target operator support. By epoch 15, exact recovery reaches \(88.0\%\) for advection--Burgers, \(89.0\%\) for advection--cubic-Burgers, \(88.0\%\) for advection--sine, and \(96.0\%\) for Burgers--cubic-Burgers. By epoch 20, all four targets are above \(90\%\). This behavior suggests that the pretrained model supplies operator-level representations that can be rapidly recombined through limited target-family fine-tuning.

The from-scratch curves in Fig.~\ref{fig:fewshot_transfer_summary}A provide a different comparison point. They show that the composite targets are learnable without pretraining, but not on the same early-epoch schedule. At 20 epochs, scratch training is still near zero exact recovery across the four composites. By 100 epochs, it reaches \(0\%\) on advection--Burgers, \(73\%\) on advection--cubic-Burgers, \(51\%\) on advection--sine, and \(100\%\) on Burgers--cubic-Burgers. The comparison measures adaptation efficiency through the required training horizon rather than hardware-dependent wall-clock time: 20 fine-tuning epochs contain 200 optimizer updates, whereas 100 scratch epochs contain 1000 updates. These values come from single training trajectories evaluated at fixed horizons, so they establish the observed optimization behavior but do not quantify variation across random initializations. Moreover, this is not an initialization-only ablation. Each fine-tuned model is adapted to one family using batches of 100, while the scratch model is trained jointly on four families using batches of 200. Under these reported protocols, pretrained checkpoint adaptation reaches high target-family recovery over a shorter optimization horizon. The result does not establish how much of the gap is attributable solely to initialization rather than to the difference between single-family and joint training.

The advection--Burgers fine-tuning run also provides the retention diagnostic in Fig.~\ref{fig:fewshot_transfer_summary}B. At each checkpoint, we compute TRR separately for each of the 20 original PDE families using fixed expected denominators, count missing outputs as failures, and report the unweighted macro-average. The corrected statistic decreases from \(96.52\%\) before fine-tuning to \(96.22\%\) at epoch 5, \(93.41\%\) at epoch 10, \(90.43\%\) at epoch 15, and \(88.50\%\) at epoch 20. It is not a pooled accuracy over all original-family samples, and it is not averaged across the other three fine-tuning experiments. The decrease is smaller than the gain on advection--Burgers, but it is monotone over this fine-tuning window. This result motivates future adaptation methods that improve new-family recovery while preserving original-family performance.

\subsection{Hierarchical discovery}
\label{experiment:hierarchical_discovery}

We evaluate the hierarchical discovery algorithm using a separate model pretrained only on advection, inviscid Burgers, inviscid cubic Burgers, and diffusion. Without updating its weights, the model is tested on \(M = 20\) trajectories from each of three unseen two-term composites: viscous Burgers, viscous cubic Burgers, and advection--diffusion. For each target equation, the symbolic form and coefficients are fixed, while the trajectories are generated from independently sampled initial conditions. We assess whether the first stage recovers the dominant operator, the residual stage identifies the subordinate term, and BFGS jointly calibrates their coefficients. The full protocol and sample-level outcomes are provided in Appendix~\ref{app:hierarchical_discovery_details}.

Table~\ref{tab:hierarchical_discovery} summarizes the recovered equations. For the two Burgers cases, the first stage identifies the appropriate nonlinear-flux operator and the residual stage recovers diffusion. The advection--diffusion case produces competing first-stage predictions, but coefficient consistency selects advection before the diffusion residual is recovered. Overall, the hierarchical discovery algorithm predicts the correct composite PDE skeleton in all three cases and yields only small coefficient errors on average after BFGS refinement.

\begin{table}[htbp]
\caption{\textbf{The hierarchical discovery algorithm recovers the tested unseen two-term composites through sequential operator identification and BFGS calibration.}}
\label{tab:hierarchical_discovery}
\centering
\small
\begin{tabular}{@{}l @{\hspace{1.5em}} l @{\;} l@{}}
\toprule
Ground truth & \multicolumn{2}{l}{Decomposition results} \\
\midrule
\raisebox{-1.5\normalbaselineskip}[0pt][0pt]{$u_t = -1.08\, uu_x + 0.00541\, u_{xx}$}
  & Step 1 (Dominant):    & $u_t = -1.14\, uu_x$ \\
  & Step 2 (Subordinate): & $u_t = 0.00476\, u_{xx}$ \\
  & Combined:             & $u_t = -1.14\, uu_x + 0.00476\, u_{xx}$ \\
  & After BFGS:           & $u_t = -1.080\, uu_x + 0.005407\, u_{xx}$ \\
\midrule
\raisebox{-1.5\normalbaselineskip}[0pt][0pt]{$u_t = -0.902\, u^2 u_x + 0.00334\, u_{xx}$}
  & Step 1 (Dominant):    & $u_t = -0.989\, u^2 u_x$ \\
  & Step 2 (Subordinate): & $u_t = 0.00505\, u_{xx}$ \\
  & Combined:             & $u_t = -0.989\, u^2 u_x + 0.00505\, u_{xx}$ \\
  & After BFGS:           & $u_t = -0.9020\, u^2 u_x + 0.003341\, u_{xx}$ \\
\midrule
\raisebox{-1.5\normalbaselineskip}[0pt][0pt]{$u_t = -0.991\, u_x + 0.00513\, u_{xx}$}
  & Step 1 (Dominant):    & $u_t = -1.025\, u_x$ \\
  & Step 2 (Subordinate): & $u_t = 0.00458\, u_{xx}$ \\
  & Combined:             & $u_t = -1.025\, u_x + 0.00458\, u_{xx}$ \\
  & After BFGS:           & $u_t = -0.9910\, u_x + 0.005131\, u_{xx}$ \\
\bottomrule
\end{tabular}
\end{table}

In real applications, the operators may have similar magnitudes or be strongly coupled, making the dominant component difficult to identify. An incorrect first-stage operator or an inaccurate neural surrogate contaminates the residual passed to the second stage, while measurement noise and discretization error can obscure weak terms. The present study is limited to clean synthetic trajectories, additive two-term PDEs, and a restricted operator vocabulary, so performance on multi-term, nonadditive, noisy, or experimental systems remains to be established.

\section{Conclusions}
We presented FoundPDE as a scoped foundation model for symbolic PDE discovery: a reusable pretrained data-to-symbol model for one-dimensional time-dependent PDEs, rather than a universal PDE solver or unrestricted open-form symbolic regression engine. The in-domain experiments show that a single transformer trained on canonical PDE families can recover symbolic structure with high accuracy. The Fourier-and-derivative model reaches \(97.50\%\) type recognition, with a coefficient-setting-level 95\% bootstrap interval of \([96.54\%,98.30\%]\), indicating that trajectory-to-equation prediction can be amortized across a broad family set. Under multiplicative corruption of the model input, recognition decreases gradually, to \(94.43\%\) at \(\eta=0.02\), and in a focused diffusion-reaction comparison FoundPDE retains most of its accuracy at noise levels at which SINDy and weak SINDy fail.

The parameter-generalization experiment is restricted to coefficient shifts within known symbolic families. On 5,120 samples drawn from coefficient supports that are strictly disjoint from the in-domain ranges, FoundPDE reaches \(90.53\%\) TRR, compared with \(97.34\%\) on the corresponding 19-family in-domain subset. The 6.81-percentage-point decrease is concentrated in sine-flux conservation and in heat, Klein--Gordon, and wave equations whose coefficients lie far beyond the training range. Advection further shows that high TRR can coexist with large relative \(L^2\) error when an extrapolated transport coefficient has the wrong sign. These results support parameter extrapolation within much of the learned symbolic vocabulary.

The few-shot fine-tuning and hierarchical discovery experiments provide evidence for extending the pretrained representation beyond its original symbolic vocabulary. Under the reported single-family adaptation protocol, a separately pretrained FoundPDE checkpoint reaches \(91.0\%\)--\(97.0\%\) exact recovery across four held-out composite families after 20 epochs, while joint four-family scratch training requires a longer optimization horizon to approach comparable behavior. Because the protocols differ in family composition and batch size, this result supports the transfer capacity of the broader FoundPDE pretraining framework over the tested training horizon rather than an initialization-only efficiency claim for the principal checkpoint. Hierarchical discovery recovers selected two-term composite examples by identifying dominant and subordinate operators separately, followed by BFGS coefficient refinement. These results indicate that the pretrained representation can support downstream extension, while systematic transfer to new symbolic compositions remains open.

Overall, FoundPDE provides evidence for pretrained symbolic PDE discovery within a canonical-family setting. Future work should extend this setting through broader transfer experiments, adaptation methods that reduce forgetting, stronger decomposition methods than the current residual pass, and extensions beyond one-dimensional synthetic PDE families toward higher-dimensional, noisier, and experimentally observed systems.

\section*{Acknowledgments}

This work was supported by the U.S. Department of Energy Office of Advanced Scientific Computing Research under Grants No.~DE-SC0025593 and No.~DE-SC0025592, and the U.S. National Science Foundation under Grants No.~DMS-2347833 and No.~DMS-2527294. This research used resources of the National Energy Research Scientific Computing Center (NERSC), a Department of Energy User Facility (project m4946).

\section*{Code availability}

The FoundPDE code will be available at \url{https://github.com/lu-group/foundpde}.

\appendix
\section{Dataset specification}
\label{app:dataset_specification}

The main corpus follows the PDE families, initial-condition generators, and numerical solvers of PROSE-PDE~\cite{sun2025foundationmodelpartialdifferential}, with modified coefficient supports. Unless otherwise specified, each positive continuous parameter \(q\) has reference value \(q_0\) and is sampled as \(q\sim\mathcal{U}(0.1q_0,1.9q_0)\). Signed and discrete parameter supports are stated explicitly in Table~\ref{tab:appendix_pde_families}, which summarizes the reference values and in-domain supports for all families. We generate \(25.6\mathrm{K}\) symbolic PDE instances spanning 20 PDE types. Each generated PDE is solved under 20 initial conditions, giving \(25.6\mathrm{K}\times20=512\mathrm{K}\) solution samples. Unless otherwise specified, the spatial domain length is \(x_f=2\); the Fokker--Planck equation follows the PROSE-PDE physical scaling with \(x_f=2\times10^{-6}\). The final time \(t_f\) depends on the PDE family and is also reported in Table~\ref{tab:appendix_pde_families}.

\begin{table}[htbp]
\caption{\textbf{The synthetic corpus contains 20 PDE types constructed from the one-dimensional benchmark families of PROSE-PDE~\cite{sun2025foundationmodelpartialdifferential}.} The table reports each family's canonical form, reference setting, and in-domain parameter support. Diffusion--reaction and conservation-law families each contain multiple symbolic types, yielding 20 PDE types in total.}
\label{tab:appendix_pde_families}
\centering
\scriptsize
\setlength{\tabcolsep}{4pt}
\begin{tabular}{@{}>{\raggedright\arraybackslash}p{0.18\linewidth}>{\raggedright\arraybackslash}p{0.29\linewidth}>{\raggedright\arraybackslash}p{0.27\linewidth}>{\raggedright\arraybackslash}p{0.19\linewidth}@{}}
\toprule
Family & Canonical form & Reference setting & In-domain support \\
\midrule
Heat / Diffusion & \(u_t = c u_{xx}\) & \(c_0=3\times10^{-3}\), \(t_f=3\) & \(c/c_0\in[0.1,1.9]\) \\
Porous medium & \(u_t = (u^m)_{xx}\) & \(t_f=0.1\) & \(m\in\{2,3,4\}\) \\
Klein--Gordon & \(u_{tt} = c^2u_{xx}-m^2c^4u\) & \(c_0=1\), \(m_0=0.1\), \(t_f=1\) & \(c/c_0,m/m_0\in[0.1,1.9]\) \\
Sine--Gordon & \(u_{tt} = u_{xx}-c\sin(u)\) & \(c_0=1\), \(t_f=1\) & \(c/c_0\in[0.1,1.9]\) \\
Cahn--Hilliard & \(u_t = -\epsilon^2u_{xxxx}-6(u u_x)_x\) & \(\epsilon_0=10^{-2}\), \(t_f=0.5\) & \(\epsilon/\epsilon_0\in[0.1,1.9]\) \\
KdV & \(u_t = -\delta^2u_{xxx}-u u_x\) & \(\delta_0=0.022\), \(t_f=1\) & \(\delta/\delta_0\in[0.1,1.9]\) \\
Advection & \(u_t = -\beta u_x\) & \(\beta_0=1\), \(t_f=3\) & \(\beta\in[-1.9,1.9]\) \\
Wave & \(u_{tt} = \beta^2 u_{xx}\) & \(\beta_0=0.5\), \(t_f=1\) & \(\beta/\beta_0\in[0.1,1.9]\) \\
Diffusion--reaction & \(u_t = \nu u_{xx}+\rho R(u)\) & \(\nu_0=3\times10^{-3}\), \(\rho_0\in\{1,0.1\}\), \(t_f=3\) & \(\nu/\nu_0,\rho/\rho_0\in[0.1,1.9]\) \\
Viscous conservation & \(u_t = -k f(u)_x+\frac{\epsilon}{\pi}u_{xx}\) & \(k_0=1\), \(\epsilon_0=10^{-2}\), \(t_f=2\) & \(k/k_0,\epsilon/\epsilon_0\in[0.1,1.9]\) \\
Inviscid conservation & \(u_t = -k f(u)_x\) & \(k_0=1\), \(t_f=2\) & \(k/k_0\in[0.1,1.9]\) \\
Fokker--Planck & \(u_t = D u_{xx}-\frac{D}{k_BT}(\nabla U(x)u)_x\) & \(\eta_0=10^{-3}\), \(t_f=0.1\), \(x_f=2\times10^{-6}\) & \(\eta/\eta_0\in[0.1,1.9]\) \\
\bottomrule
\end{tabular}
\end{table}

For the diffusion--reaction family, \(R(u)\) is one of \(u(1-u)\), \(u\), \(u(1-u)(u-a)\) with \(a\in(0,1)\), or \(u^2(1-u)^2\). For the viscous conservation laws, \(f(u)\) is one of \(\frac{1}{2}u^2\), \(u\), \(\frac{1}{3}u^3\), or \(\sin(u)\). For the inviscid conservation laws, \(f(u)\) is one of \(\frac{1}{2}u^2\), \(\frac{1}{3}u^3\), or \(\sin(u)\). These grouped choices account for the 20 PDE types used in the main corpus.

\section{Model, training, and augmentation details}
\label{app:model_training_details}

The model used in the main experiments is a transformer sequence-to-sequence network~\cite{vaswani2017attention}. The encoder uses a 512-dimensional embedding, 8 attention heads, and 4 transformer layers. The symbolic decoder uses 8 attention heads and 8 transformer layers. Input channels are first projected into the latent space by a feed-forward embedder with GELU activation, and reversible instance normalization~\cite{kim2022revin} is applied to reduce scale variation across PDE families.

The model is trained with cross-entropy loss on the target token sequence. Unless otherwise specified, we use Adam~\cite{kingma2015adam} with learning rate \(10^{-4}\), an inverse-square-root schedule with 8,000 warm-up steps, weight decay \(10^{-4}\), batch size 512, gradient clipping at 1.0, 100 epochs, and 1,000 iterations per epoch, so that each of the 512,000 training samples is processed about 100 times. Decoding is greedy in the current experiments. All four feature variants in Fig.~\ref{fig:in_domain_feature_training_combo} use this schedule. Their training times are projected from the measured time per optimization step: each variant was trained for 600 steps on a workstation with an NVIDIA GeForce RTX 3090 GPU and an AMD Ryzen Threadripper PRO 5955WX CPU with no other jobs running, and the mean time per step over steps 100--600 was multiplied by the 100,000 steps of a full run. Appendix~\ref{app:convergence} examines convergence and the effect of the training budget.

For Fourier augmentation, we compute the discrete Fourier transform along the spatial dimension and concatenate the real and imaginary parts with the original solution channel. For derivative augmentation, we compute only the first-order spatial derivative \(u_x\) using finite differences along the spatial axis. The derivative augmentation does not include \(u_t\) or \(u_{xx}\). The four main model variants are therefore: raw \(u\), \(u\) with Fourier channels, \(u\) with \(u_x\), and \(u\) with both \(u_x\) and Fourier real/imaginary channels.

For the robustness-to-noise benchmark, Gaussian corruption is pre-injected after construction of the augmented input tensor. The stored corrupted tensor therefore contains perturbed solution, first-order spatial-derivative, Fourier-real, and Fourier-imaginary channels; the derived channels are not recomputed from one corrupted raw solution field. Evaluator-side noise is set to zero during inference, so \(\eta\in\{0,0.02,0.05,0.1\}\) identifies only the stored corruption: each entry \(z\) of the augmented tensor is replaced by \(z+\eta|z|\xi\) with an independent \(\xi\sim\mathcal{N}(0,1)\). The noise is therefore multiplicative, and entries that are exactly zero remain unchanged. For relative \(L^2\) evaluation, every decoded equation is subsequently solved from the corresponding clean initial condition and compared with the noise-free trajectory; the corrupted tensor is used only as the model input.

\section{Training budget and convergence}
\label{app:convergence}

\begin{figure}[htbp]
\centering
\includegraphics[width=\linewidth]{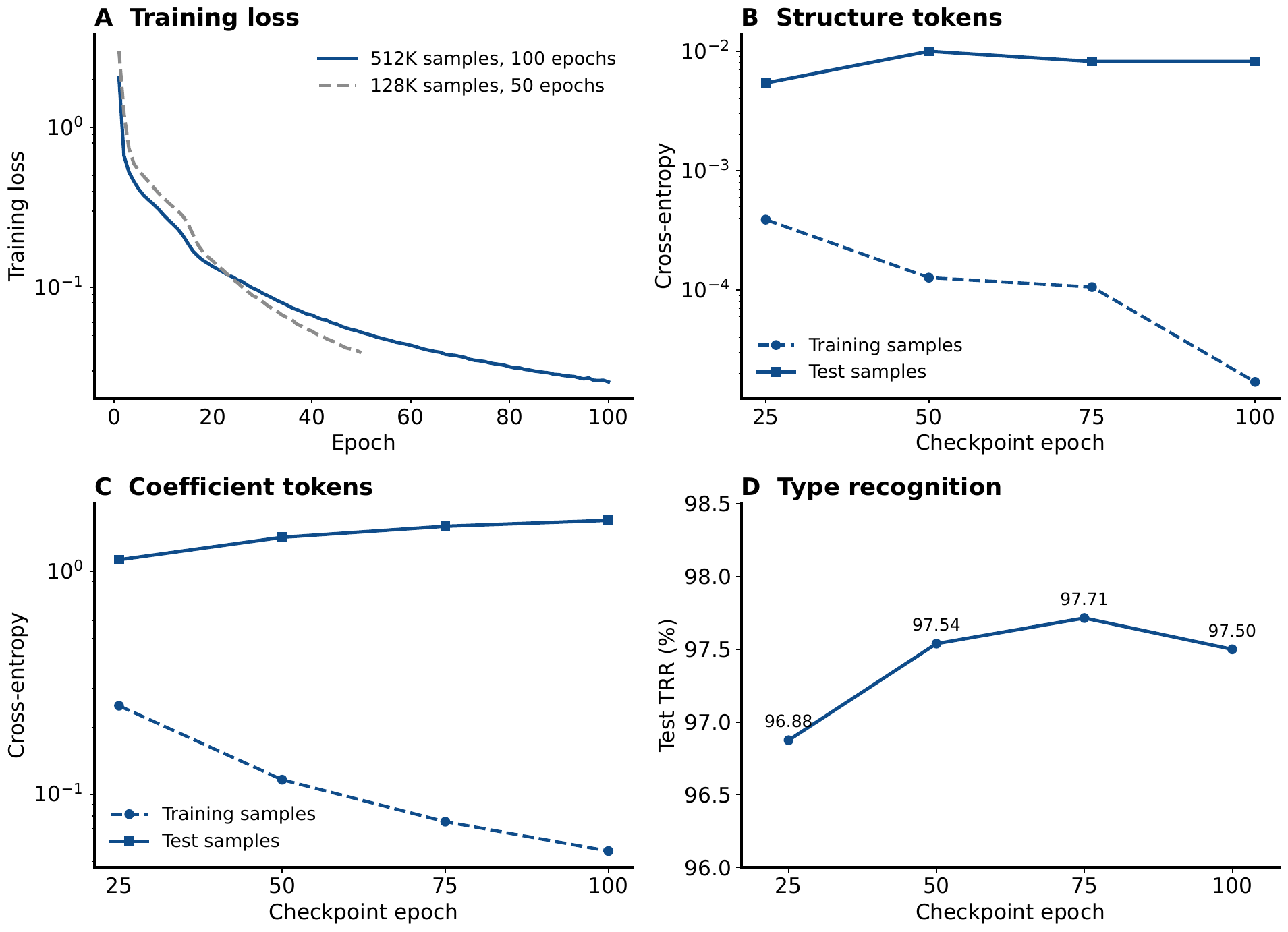}
\caption{\textbf{The principal model converges with respect to symbolic recovery before its training loss stops decreasing.} (\textbf{A}) Per-epoch mean training loss of the principal model (512,000 samples, 100 epochs) and of the same model trained on 128,000 samples for 50 epochs. (\textbf{B}, \textbf{C}) Teacher-forced cross-entropy of the principal model at each saved checkpoint on 5,120 training samples and on the 5,120 test samples, for equation-structure tokens and for coefficient tokens (sign, mantissa, and exponent). (\textbf{D}) Test TRR of each checkpoint.}
\label{fig:training_convergence}
\end{figure}

The principal model is trained for 100 epochs of 1,000 optimization steps on all 512,000 training samples. Fig.~\ref{fig:training_convergence}A shows that its training loss is still decreasing slowly at the end of training. To check whether further training would change the reported results, we evaluated the checkpoints saved every 25 epochs on the 5,120 test samples. The teacher-forced loss on equation-structure tokens (operators, variables, and derivative terms) stays between \(0.005\) and \(0.010\) at every checkpoint, with test structure-token accuracy of \(99.9\%\) (Fig.~\ref{fig:training_convergence}B), and the test TRR changes by at most \(0.21\) percentage points after epoch 50 (Fig.~\ref{fig:training_convergence}D). The continued decrease in training loss comes from coefficient tokens: training-sample accuracy on coefficient tokens rises from \(92.7\%\) at epoch 25 to \(98.2\%\) at epoch 100, whereas test accuracy remains near \(83\%\) and the test cross-entropy increases (Fig.~\ref{fig:training_convergence}C). The 512,000 samples contain 25,600 distinct equations, each solved from 20 initial conditions, so later epochs mainly fit the coefficient values of the training equations. We therefore regard the principal model as converged with respect to symbolic discovery. The epoch-100 checkpoint was fixed before the intermediate checkpoints were evaluated.

Table~\ref{tab:training_budget} compares the principal model with the same architecture and input features trained on 128,000 samples for 50 epochs with batch size 256, which processes 6.4 million samples instead of 51.2 million. The larger budget improves in-domain, corruption, and strict parameter-OOD recognition, but the gain is not uniform across strict parameter-OOD families: the reaction families improve substantially, whereas heat, Klein--Gordon, and wave are recognized less reliably when their coefficients lie far beyond the training range. In this setting, the additional training sharpens recognition within the training support at some cost to extrapolation in coefficient magnitude for these families.

\begin{table}[htbp]
\centering
\small
\caption{\textbf{A larger training budget improves recognition overall but changes the parameter-OOD failure profile.} Both models use the Fourier-and-derivative input and the same architecture. Values are TRR (\%) with the complete 5,120-sample denominators; the per-family rows are strict parameter-OOD results.}
\label{tab:training_budget}
\begin{tabular}{@{}lcc@{}}
\toprule
Setting & 128K samples, 50 epochs & 512K samples, 100 epochs \\
\midrule
Clean in-domain & 93.40 & 97.50 \\
Corruption, $\eta=0.02$ & 89.86 & 94.43 \\
Corruption, $\eta=0.1$ & 49.26 & 52.09 \\
Strict parameter OOD & 89.36 & 90.53 \\
\quad Heat & 98.53 & 67.28 \\
\quad Klein--Gordon & 85.23 & 71.97 \\
\quad Wave & 83.71 & 76.14 \\
\quad Diffusion--reaction, bistable & 65.81 & 89.34 \\
\quad Diffusion--reaction, square-logistic & 60.66 & 88.97 \\
\bottomrule
\end{tabular}
\end{table}

\section{BFGS implementation}
\label{app:bfgs_details}

BFGS~\cite{nocedal2006numerical} is applied only after the symbolic expression has been decoded. The optimizer receives the predicted expression skeleton and initial decoded coefficient values, then optimizes the floating-point coefficients against a residual loss computed from the observed trajectory and estimated derivatives. The temporal derivative follows the decoded target: the evaluator uses \(u_t\) for first-order families and \(u_{tt}\) for wave, Klein--Gordon, and Sine--Gordon. If the decoded expression cannot be parsed, if the required derivative quantities are unsupported by the evaluator, or if reconstruction fails numerically, the sample is excluded from MRCSE or relative \(L^2\) aggregation as appropriate. This is why MRCSE and relative \(L^2\) error may have different evaluable sample counts from TRR.

\section{Custom composite-family setup}
\label{app:custom_composite_setup}

The few-shot fine-tuning and from-scratch experiments use custom composite equations built on top of the PROSE-PDE data-generation infrastructure. Standard fine-tuning and from-scratch runs sample around \(\epsilon=0.01\), \(k=1\), and \(\alpha=1\) with relative range parameter \(\gamma=0.1\). Thus,
\[
\epsilon\sim\mathrm{Unif}(0.009,0.011),\qquad
\nu=\epsilon/\pi,\qquad
k\sim\mathrm{Unif}(0.9,1.1),\qquad
\alpha\sim\mathrm{Unif}(0.9,1.1),
\]
so \(\nu\in[0.002865,0.003501]\) up to rounding. The standard composite runs do not use additional coefficient clipping beyond the legacy CFL stepping, diffusion CFL control, and a final sample-norm filter \(10<\|u\|<2000\). The broader clipped ranges used in the parameter-OOD setting are separate from the standard few-shot and from-scratch experiments.

\begin{table}[htbp]
\caption{\textbf{Custom composite PDE families used for few-shot fine-tuning.} These equations combine transport, nonlinear flux, and diffusion terms that are individually familiar from pretraining but not present as full composite symbolic forms. The initial conditions for these composite datasets use a sinusoidal superposition generator rather than the Gaussian-distribution or Gaussian-process initial conditions used by some PROSE-PDE families.}
\label{tab:custom_composite_families}
\centering
\small
\begin{tabular}{ll}
\toprule
Composite family & Symbolic form \\
\midrule
Advection--Burgers & \(u_t = -\alpha u_x-k u u_x+\nu u_{xx}\) \\
Advection--cubic Burgers & \(u_t = -\alpha u_x-k u^2u_x+\nu u_{xx}\) \\
Advection--sine & \(u_t = -\alpha u_x-k\cos(u)u_x+\nu u_{xx}\) \\
Burgers--cubic Burgers & \(u_t = -\alpha u u_x-k u^2u_x+\nu u_{xx}\) \\
\bottomrule
\end{tabular}
\end{table}

The few-shot fine-tuning experiments evaluate the four composite targets in Table~\ref{tab:custom_composite_families}. We use a FoundPDE model pretrained for 100 epochs on 512,000 samples from the original PDE corpus. It shares the training corpus and schedule of the principal Fourier-and-derivative model used for the in-domain, corruption, and parameter-OOD results but takes only the solution field as input. Fine-tuning epoch 0 denotes this loaded source state before any target-family update, and a separate copy is adapted to each target family. Advection--Burgers uses 1000 training samples, corresponding to 50 coefficient settings with 20 initial conditions per setting. Each of the other three fine-tuning runs uses 500 training samples, corresponding to 25 coefficient settings with 20 initial conditions per setting. The evaluation contains 100 expected test samples for each family, corresponding to 5 coefficient settings with 20 initial conditions per setting. The from-scratch experiment trains one model jointly on all four families using 2000 training and 400 test samples, or 500 training and 100 test samples per family. For advection--cubic-Burgers, advection--sine, and Burgers--cubic-Burgers, the fine-tuning files are exact family-wise subsets of the mixed dataset used by the scratch run. The earlier advection--Burgers result uses a separately generated dataset and therefore does not provide this exact sample-level match.

Both regimes use the same optimizer family and learning rate: Adam with inverse-square-root scheduling, initial learning rate \(10^{-4}\), and 10 optimizer updates per epoch. Fine-tuning uses batches of 100 examples from one target family and runs for 20 epochs, whereas scratch training uses batches of 200 examples drawn from the mixed four-family dataset and is evaluated over a longer horizon. Thus, 20 fine-tuning epochs correspond to 200 optimizer updates and 100 scratch epochs correspond to 1000 updates. The reported trajectories are single runs rather than averages over repeated random seeds. Epochs 20 and 100 are fixed comparison horizons from these trajectories, not checkpoints selected retrospectively for their best test performance. Because family composition and batch size differ, this experiment compares the reported adaptation and scratch protocols rather than isolating initialization under otherwise identical optimization conditions.

Original-family retention is evaluated only for the advection--Burgers fine-tuning trajectory. The evaluation set contains 5,120 expected samples distributed across the 20 pretraining families. At each reported checkpoint, TRR is computed separately using each family's fixed expected count; missing or unparsable outputs are assigned zero credit. The 20 family-level TRRs are then averaged with equal family weights. The reported retention statistic is therefore a fixed-denominator macro-average across PDE families, not a pooled accuracy over all samples. The number of missing outputs increases from 2 before fine-tuning to 81 at epoch 20, and these failures are included in the corrected values.

For these composite datasets, both train and test samples use the same initial-condition family: a sinusoidal superposition generator with random modes, flips/windows, detrending, and normalization. Train and test sets differ by sampled examples and coefficient settings, not by switching to a Gaussian-process or Gaussian-distribution initial-condition family.

\paragraph{Hierarchical discovery data.}
The hierarchical discovery examples use 20 solution datasets from a fixed symbolic equation and fixed coefficients while varying initial conditions. The verified Burgers, cubic Burgers, and advection--diffusion examples use sinusoidal initial conditions and the numerical solvers.

\section{Details of the hierarchical discovery experiments}
\label{app:hierarchical_discovery_details}

\paragraph{Restricted pretraining scope and unseen targets.}
To isolate the sequential-recovery mechanism, this experiment uses a separate FoundPDE model pretrained only on advection, inviscid Burgers, inviscid cubic Burgers, and diffusion. We then evaluate three additive two-operator equations whose individual operators occur in this restricted corpus but whose complete symbolic forms are absent from pretraining:
\[
\begin{aligned}
u_t &= -0.991\,u_x + 0.00513\,u_{xx},\\
u_t &= -1.08\,u u_x + 0.00541\,u_{xx},\\
u_t &= -0.902\,u^2u_x + 0.00334\,u_{xx}.
\end{aligned}
\]
These equations correspond to advection--diffusion, viscous Burgers, and viscous cubic Burgers, respectively. The restricted setting is designed to identify one familiar operator at a time; it is not a comprehensive compositional OOD benchmark.

\paragraph{Trajectory data and residual surrogate.}
For each composite equation, we use \(M=20\) solution datasets with a fixed symbolic equation and fixed coefficients while varying the initial condition. The datasets use the sinusoidal initial-condition generator and the numerical solvers. Each trajectory contains \(N_t=32\) snapshots on a spatial grid of \(N_x=64\). The per-target surrogate \(\mathcal{N}\) is a Fourier Neural Operator~\cite{li2021fourierneuraloperatorparametric} trained for next-timestep prediction \(u(\cdot,t)\mapsto u(\cdot,t+\Delta t)\) on the same 20 trajectories supplied to FoundPDE. Thus, each target provides \(M(N_t-1)=20\times31=620\) input--output pairs for surrogate training. The trained surrogate supplies the full-dynamics update used in the residual construction of Algorithm~\ref{alg:hierarchical_discovery}.

\paragraph{First-stage aggregation and ambiguity handling.}
FoundPDE is applied separately to all 20 trajectories, producing a candidate operator type and coefficient for each trajectory. When one type clearly dominates the predictions, that type is retained and its predicted coefficients are averaged. When the votes are split, we compare the coefficient consistency of the competing candidates using the coefficient of variation, \(\mathrm{CV}=\sigma/|\mu|\), computed from the predicted coefficients assigned to each candidate. The candidate with the lower CV is selected because its coefficient estimates are more stable across initial conditions. Across the selected examples, a CV near 0.2 separates the retained assignment from the competing assignment, but this value is empirical and is not treated as a universal threshold.

\paragraph{Second-stage recovery and calibration.}
After the dominant operator is selected, its numerical evolution is removed from the surrogate full-dynamics update at every time step to construct the residual trajectories. The fixed FoundPDE model is then applied to these residual trajectories, and the second-stage predictions are aggregated to recover the subordinate diffusion operator. Finally, the two recovered operators are combined and all coefficients are jointly refined with BFGS against the original solution data.

\paragraph{Per-equation recovery outcomes.}
\emph{Viscous Burgers.} For \(u_t=-1.08\,u u_x+0.00541\,u_{xx}\), 15 of the 20 first-stage predictions identify inviscid Burgers. Averaging their coefficients gives \(u_t=-1.14\,u u_x\). In the residual stage, 13 of 20 predictions identify diffusion and give \(u_t=0.00476\,u_{xx}\). The combined equation is \(u_t=-1.14\,u u_x+0.00476\,u_{xx}\), and BFGS refines it to \(u_t=-1.080\,u u_x+0.005407\,u_{xx}\).

\emph{Viscous cubic Burgers.} For \(u_t=-0.902\,u^2u_x+0.00334\,u_{xx}\), 19 of the 20 first-stage predictions identify inviscid cubic Burgers, giving the averaged estimate \(u_t=-0.989\,u^2u_x\). In the residual stage, 18 of 20 predictions identify diffusion and give \(u_t=0.00505\,u_{xx}\). BFGS refines the combined equation to \(u_t=-0.9020\,u^2u_x+0.003341\,u_{xx}\).

\emph{Advection--diffusion.} For \(u_t=-0.991\,u_x+0.00513\,u_{xx}\), the first-stage vote is ambiguous: 12 trajectories are classified as Burgers and 8 as advection. Their coefficient variations are 0.359 and 0.037, respectively, so the consistency criterion selects advection and gives the averaged estimate \(u_t=-1.025\,u_x\). The residual stage identifies diffusion with \(u_t=0.00458\,u_{xx}\); BFGS then refines the combined equation to \(u_t=-0.9910\,u_x+0.005131\,u_{xx}\).

These stage-wise results are summarized in Table~\ref{tab:hierarchical_discovery}.

\section{Initial conditions and solvers}
\label{app:initial_conditions}

The initial-condition families are adapted from PROSE-PDE~\cite{sun2025foundationmodelpartialdifferential}. The default initial condition is a superposition of sinusoidal waves,
\[
u_0(x)=\sum_{i=1}^{N} A_i\sin(k_i x+\phi_i),
\qquad
k_i=\frac{2\pi n_i}{L_x}.
\]
Here \(n_i\) is sampled from \(\{1,\ldots,n_{\max}\}\). The amplitude and phase satisfy
\[
A_i\sim \mathrm{Unif}(0,1),
\qquad
\phi_i\sim\mathrm{Unif}(0,2\pi).
\]
Gaussian-distribution and quadratic initial conditions are retained for the equation classes listed in Table~\ref{tab:appendix_initial_conditions}. Within each family, training and in-domain testing use independent random draws from the same initial-condition generator; the split does not switch to a different initial-condition family. Table~\ref{tab:appendix_solvers} lists the numerical solver used for each equation type.

\begin{table}[htbp]
\caption{\textbf{FoundPDE uses family-specific initial-condition generators adapted from the PROSE-PDE benchmark design~\cite{sun2025foundationmodelpartialdifferential}.} Most equation families use sinusoidal superpositions, advection, wave, and Fokker--Planck use Gaussian-distribution initial conditions, and porous-medium equations use quadratic initial conditions. Training and in-domain testing draw independently from the same generator within each family.}
\label{tab:appendix_initial_conditions}
\centering
\small
\begin{tabular}{>{\arraybackslash}p{0.3\linewidth}>{\arraybackslash}p{0.31\linewidth}>{\arraybackslash}p{0.31\linewidth}}
\toprule
Equation type & Training initial condition & In-domain test initial condition \\
\midrule
\begin{tabular}[t]{@{}l@{}}
Heat, Diffusion--reaction, \\
Klein--Gordon, Sine--Gordon, \\
Cahn--Hilliard, \\
Viscous conservation, \\
Inviscid conservation, KdV
\end{tabular}
& Sinusoidal superposition with \(n_{\max}=2\) & Independent draw from the same generator \\
\midrule
Advection, wave & Gaussian distribution with \(N=2\) & Independent draw from the same generator \\
\midrule
Fokker--Planck & Gaussian distribution with \(N=1\) & Independent draw from the same generator \\
\midrule
Porous medium & Quadratic function & Independent draw from the same generator \\
\bottomrule
\end{tabular}
\end{table}

\begin{table}[htbp]
\caption{\textbf{Numerical solvers used in our dataset follow PROSE-PDE~\cite{sun2025foundationmodelpartialdifferential}.} Diffusion--reaction and conservation-law systems use the PDEBench pipeline~\cite{takamoto2022pdebench}. KdV uses a Fourier spectral solver. Fokker--Planck uses a matrix numerical method. Advection and wave equations use exact solutions defined by the initial condition. The remaining families use a method-of-lines solver.}
\label{tab:appendix_solvers}
\centering
\small
\begin{tabular}{ll}
\toprule
Equation type & Solver \\
\midrule
Heat, Klein--Gordon, Sine--Gordon, Porous medium, Cahn--Hilliard
& Method of lines \\
\midrule
Diffusion--reaction, Viscous conservation, Inviscid conservation
& PDEBench solver pipeline \\
\midrule
Advection, Wave
& Analytical solution \\
\midrule
KdV & Fourier spectral method \\
\midrule
Fokker--Planck & Matrix numerical method \\
\bottomrule
\end{tabular}
\end{table}

\section{Coefficient-setting-level uncertainty intervals}
\label{app:cluster_bootstrap}

Table~\ref{tab:cluster_bootstrap_intervals} reports 95\% percentile cluster-bootstrap intervals for the principal fixed-checkpoint evaluations. The resampling unit is a completed coefficient-instantiated equation, so all trajectories generated from its initial conditions remain together in each bootstrap replicate. The clean in-domain and corruption evaluations contain 256 equation settings with 20 initial conditions per setting; the strict parameter-OOD evaluation contains 640 settings with eight initial conditions per setting. We use 10,000 bootstrap replicates. Missing decoder outputs count as TRR failures, while relative \(L^2\) error and MRCSE are computed only on their respective evaluable prediction sets. Corruption-level relative \(L^2\) uses the clean-reference reconstruction protocol described above. The clean in-domain evaluation and the \(\eta=0\) corruption control use the same inputs and are reported as a single row. Relative \(L^2\) error excludes four predictions whose numerical solution does not terminate: the same sample at \(\eta=0.05\) and \(\eta=0.1\), and two strict parameter-OOD samples; their TRR and MRCSE contributions are unaffected. These intervals measure finite-test-set variation across sampled equations conditional on one pretrained checkpoint. They do not include training-seed, checkpoint-selection, or fine-tuning variability.

\begin{table}[htbp]
\centering
\scriptsize
\caption{\textbf{Clustered uncertainty intervals separate robust aggregate trends from heavy-tailed coefficient errors.} Values in brackets are coefficient-setting-level 95\% bootstrap intervals. TRR uses the complete 5,120-example denominator; mean relative \(L^2\) error uses reconstructable predictions; and MRCSE uses type-matched predictions with extractable packed coefficients. The column \(n_{L^2}/n_c\) reports these two evaluable counts. The widening mean-MRCSE intervals under stronger corruption and strict parameter OOD reflect a small number of extreme coefficient errors, whereas median MRCSE remains substantially more stable.}
\label{tab:cluster_bootstrap_intervals}
\resizebox{\textwidth}{!}{%
\begin{tabular}{@{}lcccccc@{}}
\toprule
Setting & Eq. settings $\times$ ICs & $n_{L^2}/n_c$ & TRR (\%) & Mean relative $L^2$ & Mean MRCSE & Median MRCSE \\
\midrule
Clean in-domain and corruption control, $\eta=0$
& $256\times20$ & $5115/4992$ & $97.50\;[96.54,98.30]$ & $0.0308\;[0.0241,0.0384]$ & $0.185\;[0.052,0.393]$ & $0.000705\;[0.000536,0.000855]$ \\
Corruption, $\eta=0.02$
& $256\times20$ & $5108/4833$ & $94.43\;[92.64,96.02]$ & $0.0358\;[0.0293,0.0435]$ & $16.87\;[0.235,49.58]$ & $0.00129\;[0.000910,0.00176]$ \\
Corruption, $\eta=0.05$
& $256\times20$ & $4868/4164$ & $81.39\;[78.44,84.10]$ & $0.0805\;[0.0700,0.0924]$ & $2150\;[10.68,5709]$ & $0.00430\;[0.00301,0.00615]$ \\
Corruption, $\eta=0.1$
& $256\times20$ & $3821/2640$ & $52.09\;[48.20,56.00]$ & $0.211\;[0.191,0.233]$ & $8.78{\times}10^{7}\;[293,2.79{\times}10^{8}]$ & $0.0206\;[0.0130,0.0302]$ \\
Strict parameter OOD
& $640\times8$ & $5011/4630$ & $90.53\;[89.22,91.80]$ & $0.156\;[0.139,0.174]$ & $0.228\;[0.140,0.331]$ & $0.00458\;[0.00395,0.00544]$ \\
\bottomrule
\end{tabular}%
}
\end{table}

\section{Parameter-OOD data}
\label{app:ood_parameter_ranges}

The strict parameter-OOD benchmark preserves the symbolic families used during pretraining while sampling coefficients outside their complete in-domain supports. For a coefficient \(q\) with reference value \(q_0\), a range ratio \([a,b]\) means \(q\in[aq_0,bq_0]\). The lower OOD boundary is set to \(1.91\), rather than the nominal in-domain endpoint \(1.9\), so that decimal serialization cannot round a generated coefficient back onto the in-domain boundary. For multi-parameter families, one coefficient is moved outside its in-domain support while the remaining coefficients are held at their reference values, except for the coupled stability restrictions listed in Table~\ref{tab:ood_parameter_ranges}. The upper limits were fixed from numerical-stability considerations and keep the generated solutions within the intended family regime. Porous-medium and structural mixed-type equations are excluded.

\begin{table}[htbp]
\centering
\scriptsize
\caption{\textbf{Strictly disjoint coefficient supports define the parameter-OOD benchmark.} Each listed interval lies outside the corresponding in-domain support in Table~\ref{tab:appendix_pde_families}. Multi-parameter equations shift one coefficient at a time unless a coupled restriction is stated. Family-specific upper limits and coupling constraints preserve numerical stability without changing the symbolic PDE family.}
\label{tab:ood_parameter_ranges}
\begin{tabular}{@{}>{\raggedright\arraybackslash}p{0.31\linewidth}>{\raggedright\arraybackslash}p{0.24\linewidth}>{\raggedright\arraybackslash}p{0.39\linewidth}@{}}
\toprule
Family & Reference coefficient(s) & Strict OOD support \\
\midrule
Heat & \(c_0=3\times10^{-3}\) & \(c/c_0\in[1.91,2.5]\) \\
Advection & \(\beta_0=1\) & \(\beta\in[-2.5,-1.91]\cup[1.91,2.5]\) \\
KdV & \(\delta_0=0.022\) & \(\delta/\delta_0\in[1.91,2.2]\) \\
Fokker--Planck & \(\eta_0=10^{-3}\) & \(\eta/\eta_0\in[1.91,2.0]\) \\
Diffusion--reaction, logistic / linear & \(\nu_0=3\times10^{-3}\), \(\rho_0\in\{1,0.1\}\) & Either \(\nu/\nu_0\in[1.91,2.2]\) or \(\rho/\rho_0\in[1.91,2.3]\) \\
Diffusion--reaction, bistable / square-logistic & \(\nu_0=3\times10^{-3}\), \(\rho_0=1\) & Either \(\nu/\nu_0\in[1.91,2.0]\) or \(\rho/\rho_0\in[1.91,2.2]\) \\
Burgers & \(k_0=1\), \(\epsilon_0=10^{-2}\) & Either \(k/k_0\in[1.91,2.2]\), with \(\epsilon\ge k/180\), or \(\epsilon/\epsilon_0\in[1.91,2.2]\) \\
Viscous conservation, linear / sine flux & \(k_0=1\), \(\epsilon_0=10^{-2}\) & Either \(k/k_0\in[1.91,2.0]\) or \(\epsilon/\epsilon_0\in[1.91,2.0]\) \\
Viscous conservation, cubic flux & \(k_0=1\), \(\epsilon_0=10^{-2}\) & \(k\in[1.91,2.0]\), \(\epsilon\in[0.012,0.019]\), and \(k\le180\epsilon\) \\
Inviscid Burgers & \(k_0=1\) & \(k/k_0\in[1.91,2.1]\) \\
Inviscid conservation, sine / cubic flux & \(k_0=1\) & \(k/k_0\in[1.91,2.0]\) \\
Cahn--Hilliard & \(\epsilon_0=10^{-2}\) & \(\epsilon/\epsilon_0\in[1.91,2.0]\) \\
Wave & \(\beta_0=0.5\) & \(\beta/\beta_0\in[1.91,2.2]\) \\
Klein--Gordon & \(c_0=1\), \(m_0=0.1\) & \(c/c_0\in[1.91,2.0]\), with \(m=m_0\) \\
Sine--Gordon & \(c_0=1\) & \(c/c_0\in[1.91,2.0]\) \\
\bottomrule
\end{tabular}
\end{table}
\section{SINDy and weak-SINDy protocol}
\label{app:sindy_protocol}

The diffusion-reaction comparison in Table~\ref{tab:reaction_results} uses one equation from each of the four reaction families. Each equation is solved from 20 initial conditions on the \(32\times64\) grid used throughout the paper, with \(t\in[0,2)\) and periodic \(x\in[0,2)\). The first initial condition is the instance used for DISCOVER. A recovered equation counts as correct when its set of active terms equals the set of monomials of the true equation.

SINDy uses the candidate library \(\{1,u,u^2,u^3,u^4,u_x,u_{xx}\}\). Temporal and spatial derivatives are computed with second-order finite differences, periodic in space, and the library columns are normalized. Sparse regression uses sequentially thresholded least squares with ridge parameter \(10^{-5}\) over 42 thresholds (zero and 41 values logarithmically spaced between \(10^{-4}\) and \(10\)). Each model is fit on the first 25 of the 32 snapshots, and the threshold is selected by the Akaike information criterion~\cite{mangan2017model} computed on the remaining seven snapshots. Weak SINDy uses the same library in weak form~\cite{messenger2021weak} with 400 rectangular test-function subdomains of half-width \(0.25\) in \(x\) and \(t\) (8 spatial and 4 temporal grid points), polynomial test functions of degree 4, and subdomain centers drawn with a fixed random seed so that every trajectory uses the same subdomains. Its threshold is selected among 50 values logarithmically spaced between \(10^{-4}\) and \(1\) by the loss of~\cite{messenger2021weak}, which balances the relative residual against the fraction of active terms.

For the noisy comparison, each observed solution value \(z\) is replaced by \(z+\eta|z|\xi\) with an independent \(\xi\sim\mathcal{N}(0,1)\). The SINDy variants receive the noisy solution. FoundPDE recomputes its derivative and Fourier channels from the same noisy solution; for reference, Table~\ref{tab:sindy_noise} also reports FoundPDE under the protocol of Sec.~\ref{sec:noise_robustness}, in which noise is injected into each channel of the clean augmented input, SINDy with Savitzky--Golay smoothing before differentiation, and weak SINDy with half-width \(0.375\). With the threshold heuristic of the PySINDy PDE-FIND example and a library that also contains products of \(u\) with its derivatives, SINDy recovers 17 of the 80 clean trajectories and selects the empty model for the linear reaction; we therefore report the AIC-based selection.

\begin{table}[htbp]
\centering
\small
\caption{\textbf{FoundPDE retains structural recovery at noise levels at which sparse-regression baselines fail.} Number of the 80 diffusion-reaction trajectories (four equations with 20 initial conditions each) whose recovered terms match the ground truth. FoundPDE (recomputed) derives its derivative and Fourier channels from the noisy solution; FoundPDE (Sec.~\ref{sec:noise_robustness}) uses noise injected into each channel of the clean augmented input. SG denotes Savitzky--Golay smoothing with the given window length, and \(H\) is the weak-form test-function half-width.}
\label{tab:sindy_noise}
\begin{tabular}{@{}lccccccc@{}}
\toprule
$\eta$ & \splitcell{FoundPDE\\(recomputed)} & \splitcell{FoundPDE\\(Sec.~\ref{sec:noise_robustness})} & SINDy & \splitcell{SINDy\\SG 11} & \splitcell{SINDy\\SG 5} & \splitcell{Weak SINDy\\$H=0.25$} & \splitcell{Weak SINDy\\$H=0.375$} \\
\midrule
0 & 68 & 68 & 50 & 9 & 27 & 53 & 50 \\
0.01 & 54 & 62 & 0 & 9 & 4 & 17 & 6 \\
0.02 & 45 & 47 & 0 & 2 & 1 & 6 & 5 \\
0.05 & 24 & 29 & 0 & 0 & 0 & 5 & 0 \\
0.1 & 5 & 23 & 0 & 0 & 0 & 1 & 2 \\
\bottomrule
\end{tabular}
\end{table}

\bibliographystyle{unsrt}
\bibliography{ref}

\end{document}